\documentclass[preprint,floatfix]{revtex4-1}

\usepackage{graphicx}
\usepackage{epsfig}
\usepackage{enumerate}
\usepackage{amssymb}
\usepackage{amsmath}
\usepackage{amsfonts}
\usepackage{dcolumn}
\usepackage{bm}
\usepackage{color}
\usepackage[colorlinks,linkcolor=blue,citecolor=blue,urlcolor=blue]{hyperref}
\usepackage[mode=text]{siunitx}

\newcommand{\be}{\begin{equation}}
\newcommand{\ee}{\end{equation}}
\newcommand{\bea}{\begin{eqnarray}}
\newcommand{\eea}{\end{eqnarray}}

\def\a{\alpha}

\def\g{\gamma}

\def\d{\delta}
\def\D{\Delta}
\def\e{\epsilon}
\def\ve{\varepsilon}

\def\l{\lambda}

\def\m{\mu}
\def\n{\nu}

\def\r{\rho}

\def\S{\Sigma}

\def\w{\omega}

\def\bla{{\mathbf a}}

\def\bld{{\mathbf d}}

\def\blk{{\mathbf k}}

\def\blq{{\mathbf q}}

\def\blU{{\mathbf U}}

\def\callP{\mbox{$\mathcal{P}$}}

\def\ua{\uparrow}
\def\da{\downarrow}

\def\bra{\langle}
\def\ket{\rangle}
\def\grad{\mbox{\boldmath $\nabla$}}

\def\1op{\hat{\mathbbm{1}}}

\def\AA{\mathring{\mathrm{A}}}

\begin{document}

\begin{center}

\textbf{\large{Exceptional Excitons}}
\vspace{0.5cm}

\vspace{0.5cm}

Enrico Perfetto$^{1,2}$ and Gianluca Stefanucci$^{1,2}$

\vspace{0.5cm}

\textit{\footnotesize{$^1$Dipartimento di Fisica, Universit{\`a} di
Roma Tor Vergata, Via della Ricerca Scientifica 1,
00133 Rome, Italy\\
$^2$INFN, Sezione di Roma Tor Vergata, Via della Ricerca Scientifica
1, 00133 Rome, Italy}}\\
\vspace{0.5cm}

%

\vspace{1cm}

\end{center}

\textbf{
Non-Hermitian physics is reshaping our understanding of quantum 
systems by revealing states and phenomena without Hermitian counterparts. 
While non-Hermiticity is typically associated with gain–loss processes in open systems,
we uncover a fundamentally different route to non-Hermitian  behavior emerging 
from  non-equilibrium correlations.
In photoexcited 
semiconductors, the effective interaction between electrons and holes
gives rise to a pseudo-Hermitian Bethe–Salpeter Hamiltonian
(PH-BSH) that governs excitonic states in the presence of excited populations.
Within this framework, we identify a previously unknown class of excitonic
quasiparticles --{\it  exceptional excitons} -- corresponding to exceptional points
embedded inside the electron–hole continuum. Exceptional excitons emerge at
the onset of population inversion, and represent 
the strongly renormalized counterparts of the system’s equilibrium excitons.
They are spatially 
localized, protected against hybridization with the continuum, and remain
long-lived even in regimes where conventional excitons undergo a Mott 
transition. Crucially, exceptional excitons appear only when the PH-BSH is
evaluated with non-thermal, 
resonantly generated carrier populations that support an excitonic superfluid.
Ab initio results for monolayer WS$_{2}$ explicitly demonstrate this scenario and 
show that exceptional excitons can be realized with 
existing ultrafast pumping techniques. We also identify distinctive
optical and photoemission signatures that enable their unambiguous detection.
}

\newpage

Over the past few years, non-Hermitian physics has 
emerged as one of the most active frontiers in modern quantum
science~\cite{Rotter_2009,Ashida02072020,NH1,RevModPhys.93.015005},
reshaping our understanding of open systems and unveiling
phenomena with no Hermitian counterparts.
Non-Hermiticity typically arises when a system is coupled to its environment,
enabling gain–loss processes or particle 
exchange~\cite{doi:10.1142/S1230161222500044,PhysRevApplied.13.014047}.
Within this framework, exceptional 
points~\cite{Heiss_2012,ep1,doi:10.1126/science.aar7709,EP2,EP3} (EP) -- singularities
where eigenvalues and eigenvectors coalesce -- play a central role.
EP give rise to a wealth of unconventional effects, including
nonreciprocal transport~\cite{nonrectra1,doi:10.1126/science.1246957}, enhanced 
sensitivity~\cite{PhysRevLett.112.203901,EPsensing1,PhysRevLett.125.240506}, and topological 
mode conversion~\cite{Uzdin_2011,conv1,PhysRevLett.118.093002}, stimulating intense efforts to exploit them for
both fundamental research and technological innovation.
Among the expanding taxonomy of exceptional 
points~\cite{PhysRevLett.123.066405,PhysRevLett.127.186602,PhysRevB.104.L201104,class1} -- classified by
their order, topology, and symmetry—a particularly intriguing 
class are the recently discovered exceptional bound states in
the continuum (EBIC)~\cite{PhysRevLett.134.103802,EBIC2}. These states represent unique non-Hermitian
singularities formed by coalesced, spatially localized modes 
that remain perfectly decoupled from the surrounding continuum,
despite residing within it 
energetically~\cite{PhysRevA.89.052132,PhysRevB.80.165125,QinShiOu+2022+4909+4917}. 
Their simultaneous long lifetimes and extreme susceptibility 
to perturbations make EBIC uniquely suited for controlling 
dissipation~\cite{xu2023recent,diss1,diss2}, mitigating 
decoherence~\cite{PhysRevLett.122.073601,Paulisch_2016}, and realizing ultrahigh quality 
factor 
platforms~\cite{highq1,PhysRevLett.119.243901,PhysRevLett.121.193903,Bezus:18}.
Yet, the experimental realization of EP
remains challenging, as it requires precisely engineered gain 
and loss channels that mediate controlled energy exchange with 
the environment.

\begin{figure}[tbp]
  \centering
  \includegraphics[width=0.99\textwidth]{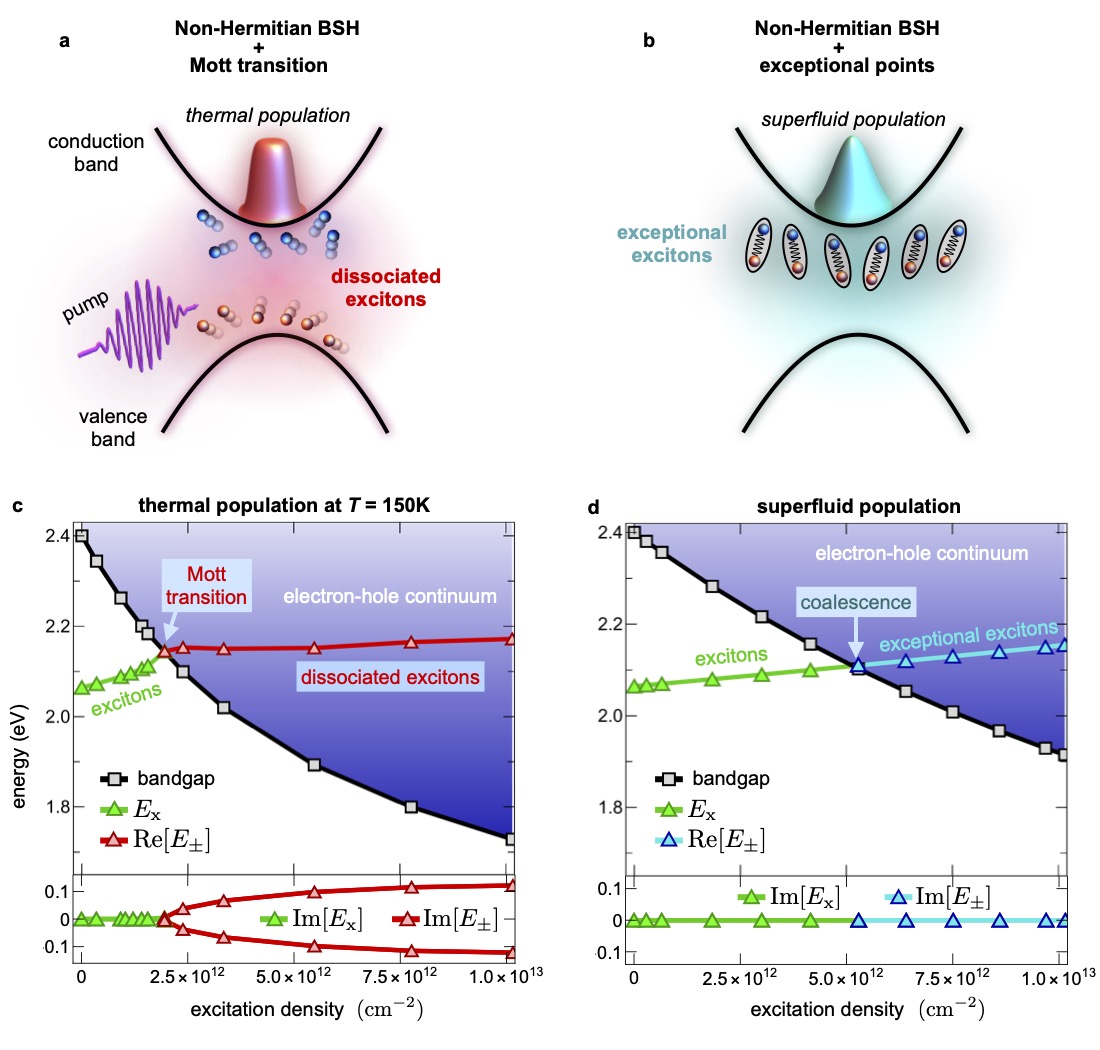}
  \caption{{\bf Excitons in photoexcited materials.}
  Panel a: Schematic illustration of exciton evaporation above the Mott 
  density in the presence of thermal carrier distribution.
  Panel b: Illustration of exciton survival in the presence of high carrier 
  population in the superfluid state. Panel c: (upper part) Evolution 
  of the real part 
  of the A-exciton 
  energy and of the bandgap (i.e. the 
  onset of the e-h continuum) for WS$_{2}$, evaluated from Eq.~(\ref{phbsh}) with thermal 
  carrier populations $f^{{\rm th}}_{\n \blk}$ at 150~K, by varying the
  excitation density $n$; (lower part) imaginary part of A-exciton 
  energy.
  Panel d: Same as panel c, but with PH-BSH evaluated from 
  Eq.~(\ref{phbsh}) with superfluid populations $f^{{\rm sf}}_{\n \blk}$ 
  obtained by solving self-consistently  
  Eq.~(\ref{eq:exciton_sf}).
  Here and in the rest of the paper the excitation 
  density $n$ is the sum of densities at the $K$ and $K'$ valleys of  
  WS$_{2}$, see Methods sections.  
  }
  \label{fig1}
  \end{figure}

In this work we introduce a fundamentally new paradigm in which non-Hermitian
Hamiltonians hosting EBIC emerge intrinsically in photoexcited
semiconductors, without requiring any external reservoirs or 
artificial gain–loss engineering. 
Within our framework, non-Hermiticity originates
directly from non-equilibrium many-body correlations~\cite{haug2009}.
This gives rise to a pseudo-Hermitian
Bethe–Salpeter Hamiltonian (PH-BSH) that governs the excitonic
spectrum of the photoexcited 
material~\cite{haug2009,Hannewald2004,KIRA2006155,perfetto_nonequilibrium_2015,PhysRevB.94.245303}.
In the conventional picture (see Fig.~\ref{fig1}a), when thermal carrier populations approach inversion,
the PH-BSH acquires complex excitonic eigenvalues, and the corresponding bound 
states disappear, signaling the excitonic Mott 
transition~\cite{PhysRevB.7.1508,haug2009,doi:10.7566/JPSJ.83.084702,PhysRevB.80.155201,Steinhoff2017}.
The key insight of the present work is that this picture changes 
qualitatively when the PH-BSH is evaluated using non-thermal carrier
distributions generated 
under intense resonant excitation—conditions that lead to the formation
of an exciton superfluid (see Fig.~\ref{fig1}b).
In this regime, we predict the emergence of EBIC as a new class of excitonic
quasiparticles, which we term {\it exceptional excitons}. 
Exceptional excitons arise once population inversion sets in and 
constitute the strongly non-equilibrium–renormalized 
counterparts of the system’s equilibrium excitons.
They originate from the coalescence of pairs of complex-conjugate eigenvalues 
into real-valued exceptional points embedded within the electron–hole continuum. 
These states  remain spatially localized, are protected from hybridization  
with continuum states, and are remarkably resilient to the excitonic Mott transition.

We explicitly demonstrate this scenario through ab initio 
calculations in  WS$_{2}$ monolayers, showing that exceptional 
excitons are potentially accessible 
with current ultrafast spectroscopic techniques. We further identify 
distinctive experimental fingerprints -- across optical and photoemission
measurements -- for their unambiguous detection, and suggest a direct
connection between eigenstate coalescence and the BEC–BCS
crossover recently observed in time-resolved ARPES on resonantly 
excited WS$_{2}$~\cite{pareek2024driving}.
Although WS$_{2}$ serves as our concrete example, the proposed underlying 
mechanism is fully general and material-independent, 
providing a universal pathway to non-Hermitian singularities in
driven quantum materials.

\bigskip
\section*{Results}
\subsection*{Exceptional Points in the PH-BSH}
The behavior of excitons is profoundly altered when a material is 
driven out of 
equilibrium~\cite{haug2009,Hannewald2004,KIRA2006155,malic1,Caruso_2026}. 
In photoexcited semiconductors, non-equilibrium carrier populations simultaneously 
renormalize band 
energies~\cite{PhysRevB.32.2266,PhysRevB.41.8288,pogna,PhysRevB.106.L081117} and weaken the effective electron–hole 
interaction~\cite{bgr1,steinhoff2,perfetto_real_2022,perfetto_real-time_2023}. 
Both effects are efficiently captured by the non-equilibrium 
Bethe–Salpeter Hamiltonian~\cite{haug2009,Hannewald2004,KIRA2006155,perfetto_nonequilibrium_2015,PhysRevB.94.245303},
defined as
\be
H_{cv \blk , c'v'\blk' }=(\e^{\rm r}_{c\blk }-\e^{\rm r}_{v \blk 
})\d_{\blk \blk '}\d_{vv'}\d_{cc'} -(f_{v \blk}-f_{c \blk}) K_{cv 
\blk , c'v'\blk' },
\label{phbsh}
\ee
where  $f_{\nu \blk}$ denotes the carrier population in the valence
($\nu=v$) and conduction ($\nu=c$) band,
$\e^{\rm r}_{\nu\blk }$ is the corresponding 
renormalized band dispersion, and
$K$ represents the Hartree plus screened-exchange (HSEX) kernel.
The eigenvalues and eigenvectors of 
$H$ provide direct access to the evolution of excitonic states under photoexcitation. 
It is worth noting that the  asymmetric Pauli-blocking factors $f_{v \blk}-f_{c \blk}$
appearing in front of the kernel $K$ play a twofold role: 
they reduce the effective electron-hole attraction and
at the same time they break Hermiticity, rendering the  
the Hamiltonian {\it pseudo-Hermitian}.
For low excitation densities, however,
it can be shown that the eigenvalues of $H$ are the same as those 
of a Hermitian matrix~\cite{PhysRevB.94.245303}, and are therefore all real.
At higher excitation densities overcoming population inversion, instead, 
the reality of eigenvalues is no longer guaranteed.

The spectral properties of the pseudo-Hermitian Bethe–Salpeter Hamiltonian (PH-BSH)
in Eq.~(\ref{phbsh}) have been 
extensively  investigated for the case of thermal carrier populations $f_{\nu 
\blk}\equiv f^{{\rm th}}_{\nu \blk}$ 
obeying Fermi-Dirac 
statistics~\cite{haug2009,PhysRevLett.84.2010,10.1063/1.5017069,doi:10.7566/JPSJ.83.084702,steinhoff2,Steinhoff2017,lohof}.
 Such distributions are expected to emerge in the incoherent regime,
typically a few picoseconds after photoexcitation.
The well-established scenario is that, as the excitation density $n$
increases, the exciton binding energy decreases while the renormalized bandgap
$E^{{\rm r}}_{g}={\rm min}[\e^{\rm r}_{c\blk }-\e^{\rm r}_{v \blk }]$
progressively narrows.
This canonical behavior is illustrated in Fig.~\ref{fig1}c which shows the spectrum of
$H$ for a WS$_{2}$ monolayer under thermal excitation  at temperature 
$T=150$~K, see also Methods section.
As expected, increasing $n$ induces a pronounced blueshift of the A-exciton resonance
$E_{\rm{x}}$ (whose equilibrium value is $E^{{\rm 
eq}}_{\rm{x}}\approx 2.1$~eV~\cite{PhysRevLett.113.076802,Hsu_2019}),
accompanied by a substantial reduction of the bandgap (whose 
equilibrium value is $E^{{\rm eq}}_{\rm{g}} \approx 2.4$~eV~\cite{PhysRevLett.113.076802,Hsu_2019}).
For densities $n\lesssim n_{c}= 2.2 \times 
10^{12}\rm{cm}^{-2}$, population inversion is not achieved, 
the  spectrum remains entirely real, and the A-exciton persists as a bound state.
At the critical density $n \approx n_{c}$ the  exciton energy $E_{\rm{x}}$ touches the 
electron-hole continuum.
Upon further incresing the carrier population,
the phase-space filling 
factors change sign for certain momenta, effectively turning the excitonic interaction
from attractive to repulsive.
As a result, the A-exciton ceases to exist as a bound-state and dissolves into the scattering continuum 
— a process known as the {\it excitonic Mott transition}.
Beyond this point, as shown in Fig.~\ref{fig1}c, the excitonic solution bifurcates
into a pair of complex-conjugate eigenvalues
$E_{\pm}$, 
corresponding to resonances embedded within the e-h continuum
and exhibiting gain or loss (absorption) features
 depending on the sign 
of ${\rm Im}E_{\pm}$.

Notice that in the above analysis we have neglected the 
renormalization of the Bethe–Salpeter kernel 
$K$ due to screening effects induced by the photoexcited carriers. 
Including these effects would  shift the 
occurrence of the Mott transition to lower excitation densities, 
precedeing the onset of population 
inversion~\cite{doi:10.7566/JPSJ.83.084702,lohof}.

 \begin{figure}[tbp]
   \centering
   \includegraphics[width=0.99\textwidth]{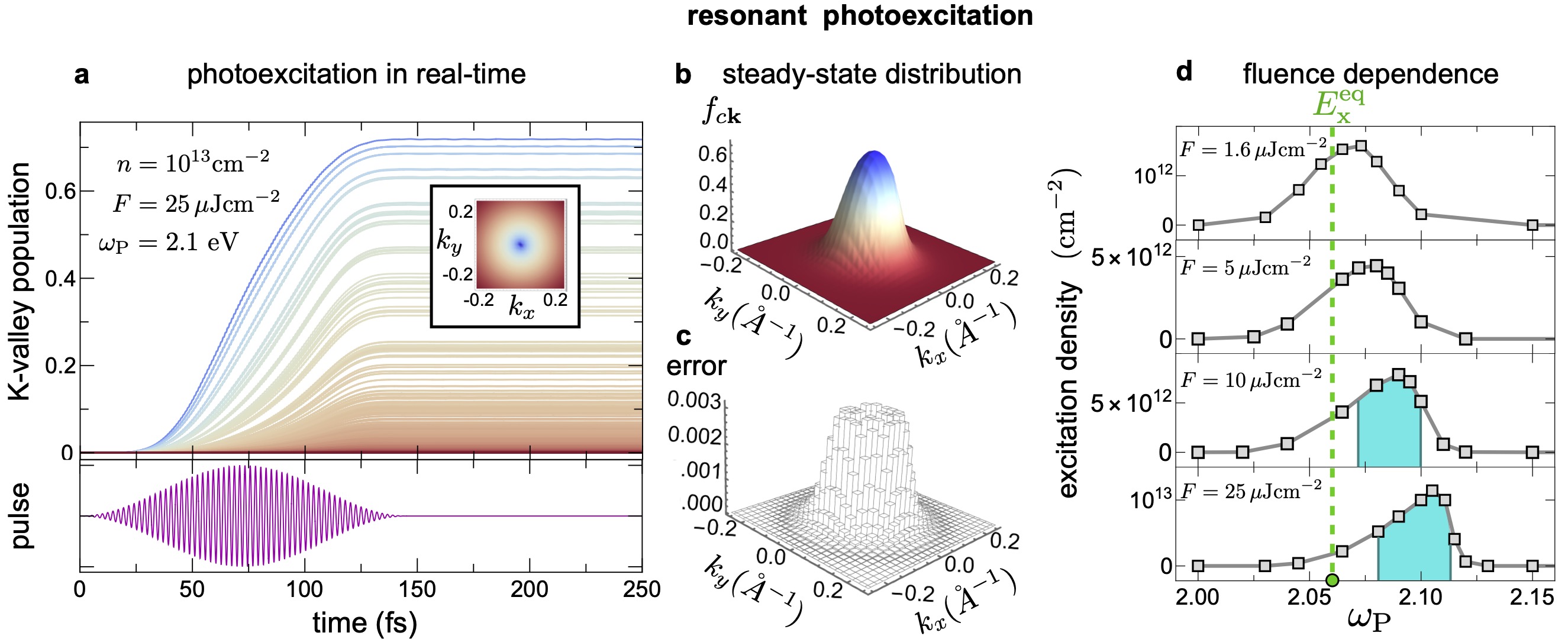}
   \caption{{\bf Real-time generation of superfluid population.}
   Panel a: (upper part) Real-time evolution of the conduction-band occupations
   $f_{c \blk}(t)$ in the $K$-valley of WS$_{2}$ following photoexcitation, obtained using time-dependent HSEX simulations
   (see Methods section). The temporal profile of the pump pulse (shown in the 
   lower part) has a duration of $\sim 100$~fs, central frequency $\omega_{{\rm P}}=2.1$~eV and 
   a fluence $F=25~\mu{\rm Jcm}^{-2}$. Occupations $f_{c \blk}(t)$
   are calculated within a  square plaquette of side  
   $0.4~\AA^{-1}$ centered at the $K$-point; the corresponding 
   colormap is shown in the inset.
   The total excited-carrier density is $n=10^{13}{\rm cm}^{-2}$. Panel b: 
   Three-dimensional visualization of the steady-state distridution 
   $f_{c \blk}$ across all momenta in the plaquette.  Panel c: 
   momentum-dependent discrepancy $|f^{{\rm sf}}_{c \blk}-f_{c \blk}|$ between the 
   superfluid distribution evaluated according to Eq.~(\ref{eq:exciton_sf})
   and the steady-state values evaluated with the real-time 
   simulation. The relative error is less than 1\%. Panel d: 
   steady-state excitation density $n$ as a function of the 
   pump frequency $\w_{{\rm P}}$ for different values of the 
   fluence. Light-blue shading denotes the parameter regime where population
   inversion occurs and the exceptional points basin is reached with the error in the 
   carrier distribution of the same order as in panel c. The equilibrium 
   value of the A-exciton energy $E^{{\rm eq}}_{{\rm x}}=2.06$~eV is 
   indicated with a vertical green line.  
   }
   \label{fig2}
 \end{figure}

We now demonstrate that such typical behavior changes dramatically when the photoexcited 
carriers follow a non-thermal distribution induced by resonant 
excitation.
Under resonant pumping, the many-body state left after the pulse is
an excitonic superfluid described by a BCS-like 
wavefunction~\cite{PSMS.2019,PhysRevLett.125.106401}.
In this regime, the carrier populations are given by
$f_{\nu \blk} = \sum_{\xi} |\varphi^{\xi}_{\nu \blk}|^2 
\equiv f_{\nu \blk}^{\mathrm{sf}} $,
where $\varphi^{\xi}_{ \nu k}$ are the lower-energy solutions 
of the self-consistent excitonic-insulator equation
(see Methods section)~\cite{PSMS.2019,PhysRevLett.125.106401}:
\be
(h^{\rm{HSEX}}_{\blk} -\mu 
)\varphi^{\xi}_{\blk}=e^{\xi}_{\blk}\varphi^{\xi}_{\blk}.
\label{eq:exciton_sf}
\ee
Here $h^{\mathrm{HSEX}}_\blk$ is the single-particle Hartree plus screened-exchange (HSEX) Hamiltonian 
evaluated at the superfluid carrier populations
$f_{\nu \blk}^{\mathrm{sf}}$, while the diagonal matrix $\mu $ 
contains different chemical potentials $\mu_{v}$ and $\mu_{c}$ for valence and conduction electrons 
respectively. The values of $\mu_c$ and $\mu_v$ depend on the pump 
fluence, and determine the excitation density $n$~\cite{PSMS.2019}.
It is worth noting that, in this regime, neglecting the screening-induced
renormalization of the Bethe–Salpeter kernel is fully justified.
The exciton superfluid described by the solution of Eq.~(\ref{eq:exciton_sf}) exhibits an
intrinsically poor screening capability~\cite{PhysRevB.102.085203}. When photoexcited electrons
and holes bind to form 
excitons—behaving as microscopic electric dipoles—their ability to 
screen the long-range Coulomb interaction is strongly  
suppressed~\cite{PhysRevB.102.085203}.

Figure~\ref{fig2}a-c shows that the superfluid distribution $f_{\nu \blk}^{\mathrm{sf}}$
obtained from Eq.~(\ref{eq:exciton_sf})
is not merely a theoretical construct but can be experimentally
realized using standard resonant laser pulses of duration $\sim 100$~fs and 
frequency
$\omega \approx E_{{\rm 
x}}^{\mathrm{eq}}$~\cite{PSMS.2019,PhysRevLett.125.106401,PhysRevMaterials.5.083803,PhysRevB.103.L241404},
as detailed in the Methods section.
\begin{figure}[tbp]
   \centering
   \includegraphics[width=0.5\textwidth]{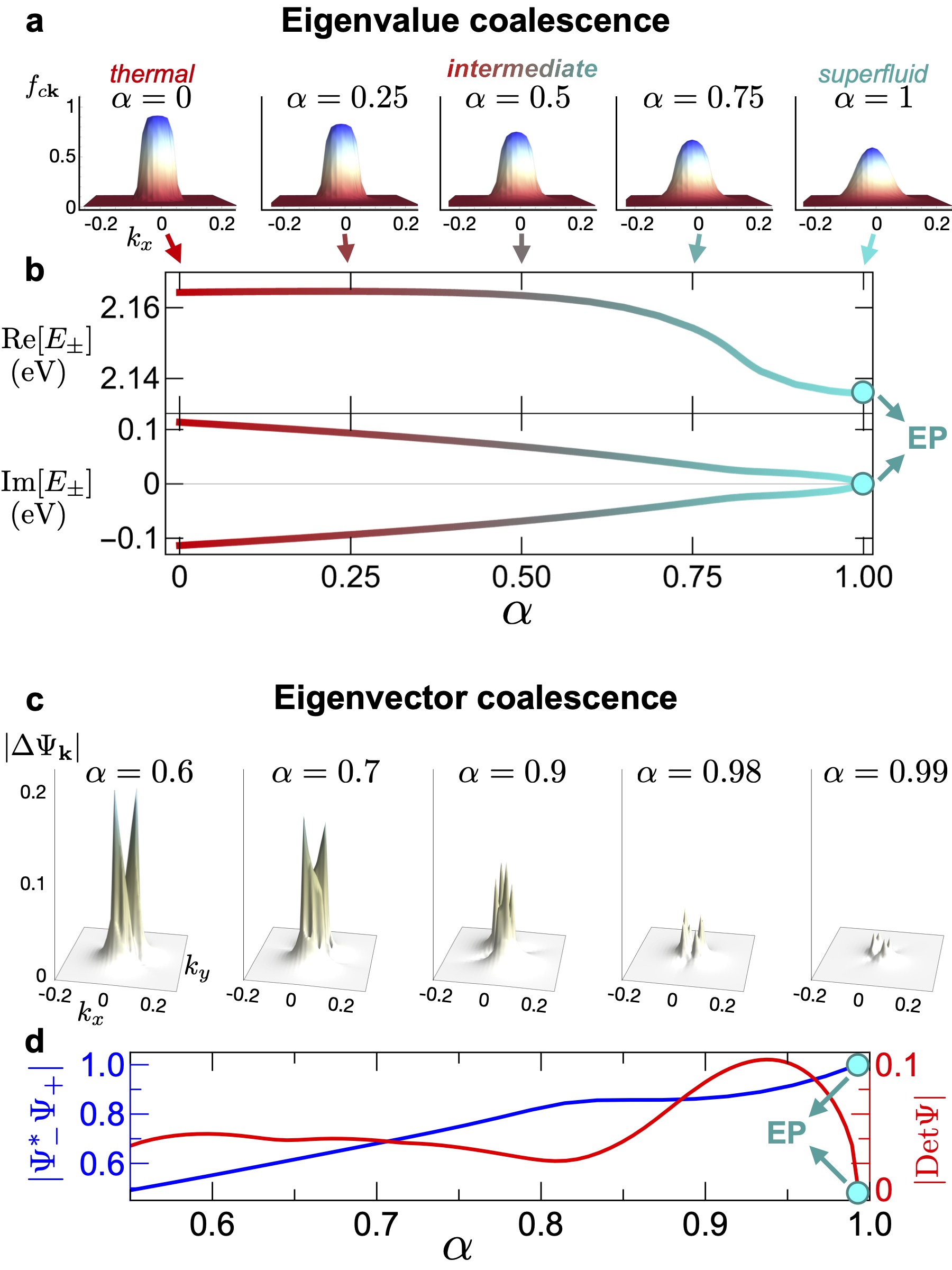}
   \caption{{\bf Excitonic eigenmode coalescence.}
   Panel a: Evolution of the conduction-band carrier distribution $f_{c \blk}$ 
   in WS$_{2}$ corresponding to a total excitation density 
   $n=7.5 \times^{12}{\rm cm}^{-2}$, as the control parameter $\a$
   is tuned from the thermal limit $f_{c \blk}= f^{{\rm th}}_{c \blk}$ 
   at $T=150~$K ($\a=0$)  to the superfluid limit
   $f_{c \blk}= f^{{\rm sf}}_{c \blk}$ ($\a=1$). Panel b: 
   Real and imaginary parts of the complex-conjugate excitonic 
   eigenvalues $E_{\pm}$ as a function of $\a$. 
   Panel c: Eigenvector similarity
   $|\D \Psi_{\blk}|$ for different  representative values of $\a$.
   Panel d: Normalized overlap $|\Psi^{\ast}_{-}\Psi_{+}|$ (blue curve) 
   and determinant of the eigenvector matrix $|{\rm Det} \Psi|$ (red curved) as a function of $\a$.
   In  panels c and d, the eigenvectors $\Psi_{\pm}$ are  normalized to 1.}
   \label{fig3}
 \end{figure}
This superfluid distribution exhibits a remarkable property---our first main finding.
We demonstrate that the BCS excitonic state
$\Psi^{{\rm sf}}_{cv\blk}=\sum_{\xi}\varphi^{\xi}_{c \blk}\varphi^{\xi *}_{v \blk}$
is a bound eigenvector of $H$ with real eigenvalue  $E_{{\rm x}} = \mu_c - 
\mu_v$ for any value of the excitation density 
(see Supplementary Note 1). 
Following a weak resonant excitation,
$\Psi^{{\rm sf}}$ coincides with the renormalized A-exciton wavefunction,
with a blue-shifted energy $E_{\rm x}  > E_{\rm 
x}^{\mathrm{eq}}$ (see Fig.~\ref{fig1}d).
As the excitation density increases, $E_{\rm x}$ approaches the renormalized bandgap 
$E_g^{{\rm r}}$, reaching $E_{\rm x} = E_g^{{\rm r}}$ at the critical
density $n_c \approx 5.3 \times 10^{12}\,\mathrm{cm^{-2}}$,
corresponding to population inversion. Beyond this point,
the exciton energy enters the particle--hole continuum, yet it remains real.
Thus, $\Psi^{{\rm sf}}$ must be intepreted as the renormalized A-exciton wavefunction in the superfluid phase, 
becoming a bound state embedded in the electron-hole continuum above the critical 
density.
A closer inspection of the PH-BSH spectrum reveals our second main finding.
For $n > n_c$, the PH-BSH becomes \textit{defective}: 
the eigenvalue $E_{{\rm x}} = \mu_c - \mu_v$ acquires algebraic multiplicity two, 
with the corresponding eigenvectors coalescing into $\Psi^{{\rm sf}}$. 
This indicates that, at high superfluid densities, the exciton transforms 
into an exceptional point.
The eigenvalue coalescence is visualized in Fig.~\ref{fig3},
where we follow the evolution of the  excitonic eigenvalues of PH-BSH 
as the carrier distribution interpolates
between the thermal and superfluid limits,
$f_{\nu \blk}=(1-\a)f^{\rm{th}}_{\nu \blk}+\a f^{\rm{sf}}_{\nu \blk}$,
at fixed excitation density $n = 7.5 \times 10^{12}\,\mathrm{cm^{-2}}$
(see Fig.~\ref{fig3}a).
Figure~\ref{fig3}b illustrates that for $\alpha = 0$ (thermal distribution above population inversion),
the pair of complex-conjugate excitonic eigenvalues $E_{\pm}$
coincide with the values reported in Fig.~\ref{fig1}a, namely
$E_{\pm} \approx 2.16 \pm i\,0.11~\mathrm{eV}$.
As $\alpha$ increases, the real part of $E_{\pm}$ 
slightly decreases, while the imaginary parts 
converge toward zero. In the superfluid limit $\alpha \to 1$, 
the two eigenvalues merge into a single real value $E_{+}=E_{-} = E_{{\rm 
x}} \approx 2.14~\mathrm{eV}$, 
signaling the emergence of a second-order exceptional point (see also 
Supplementary Note 2).
The squareroot-like behavior of ${\rm Im[E_{\pm}]}$ in the vicinity
of $\a =1$ reflects the non trivial topological properties 
of this exceptional point, as discussed in the Supplementary Note 3.
The collapse of the corresponding eigenvectors $\Psi_{+}$ and 
$\Psi_{-}$ is illustrated in 
Fig.~\ref{fig3}c,d, where we evaluate three independent coalescence
indicators, namely the eigenvector similarity
$\D \Psi_{\blk}\equiv |\Psi_{+\blk}-\Psi_{-\blk}|$,
the normalized overlap 
$|\Psi^{\ast}_{-}\Psi_{+}|\equiv 
|\sum_{\blk}\Psi^{\ast}_{-\blk}\Psi_{+\blk}|$,
and the determinant of the eigenvector matrix $|{\rm Det} \Psi|$.
As the control parameter $\a$ is increased from 0.6 to 1
all the three
quantities exhibit a clear and systematic convergence of $\Psi_{+}$ and 
$\Psi_{-}$: specifically $\D \Psi_{\blk} \to 0$,  
$|\Psi^{\ast}_{-}\Psi_{+}| \to 1$ and $|{\rm Det} \Psi| \to 0$.
Taken together, these signatures provide unambiguous evidence for the coalescence  of  $\Psi_{+}$ and 
$\Psi_{-}$ into the exceptional eigenvector $\Psi^{{\rm sf}}$,
demonstrating that $H$ becomes defective at this point.
These findings converge toward our central result: 
the exceptional point emerging in the PH-BSH spectrum at high 
superfluid carrier density corresponds to an excitonic EBIC
-- a state we term the {\it exceptional exciton}. 

A brief comment on the experimental realization of this state is in order. 
The pump frequency required to enter a basin of carrier distributions
that are close enough to generate EP 
does not coincide with the equilibrium exciton energy $E^{{\rm eq}}_{{{\rm 
x}}}$, but is consistently higher, see  Fig.~\ref{fig2}d.
The remainder of the paper is devoted to a detailed
characterization of the exceptional exciton. In particular,
we demonstrate how it remains a bound state embedded in the continuum,
and we discuss the distinctive signatures it imprints 
on optical and ARPES spectra.

\begin{figure}[tbp!]
  \centering
  \includegraphics[width=0.5\textwidth]{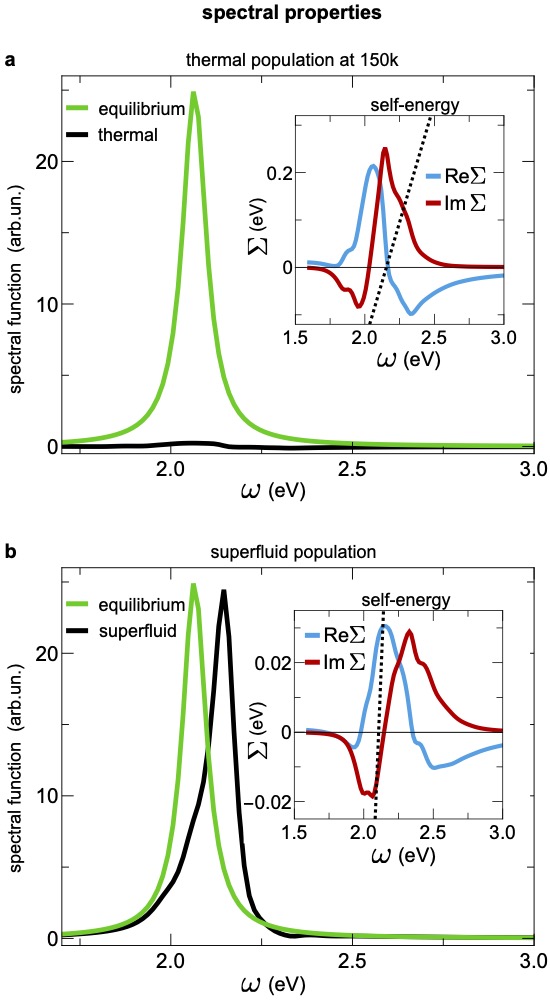}
  \caption{{\bf Spectral properties of excitonic modes.}
  Panel a: Spectral function $A_{{\rm x}}(\w)$ defined in Eq.~(\ref{spec}) 
  for WS$_{2}$ with an excitation density $n=10^{13}{\rm cm}^{-2}$ in the presence of thermal carriers
  at $T=150$~K (black curve). For comparison, the corresponding equilibrium spectral 
  function obtained by setting $H=H^{{\rm eq}}$ (i.e. $\d H=0$)  is also shown (green curve).  
  The inset displays the real (blue curve) and imaginary (red curve) parts of the embedding 
  self-energy $\S(\w)$, together with the line $\omega -H_{{\rm xx}}$ 
  (dashed). Panel b: Same quantities as in panel a,
  but evaluated for superfluid carriers at the same excitation density $n=10^{13}{\rm cm}^{-2}$.
  In both panels the equilibrium spectral function  
  and the self-energy are computed using $\eta=40$~meV.
  }
  \label{fig4}
  \end{figure}

\subsection*{Spectral Properties}
To determine whether the A-exciton, once transformed into
an exceptional point, survives as a bound state 
within the e-h band,
we employ the resolvent method, which is frequently
used in the study of bound states in the continuum.
Let us consider the  equilibrium Bethe-Salpeter equation
$H^{{\rm eq}}\Psi^{{\rm eq}}_{\l}=E^{{\rm eq}}_{{\rm \l}}\Psi^{{\rm 
eq}}_{\l} $, where $H^{{\rm eq}}$ is the (Hermitian) matrix in 
Eq.{\ref{phbsh}} at $f_{c\blk}=0$ and $f_{v\blk}=1$, and where the index 
$\l={\rm{x}}$ denotes the 
equilibrium A-exciton state. We define the mixing Hamiltonian $\d 
H=H-H^{{\rm eq}}$ which acts as an effective non-Hermitian coupling term encoding the gain
and loss channels associated with decay processes. 
This coupling drives the hybridization of the A-exciton with the 
continuum of states of $H$,
that emerge in the presence of excited carriers.
The extent of this hybridization can be quantified by 
examining the singular behavior of the spectral function
\be
A_{{\rm x}}(\omega)=-{\rm Im}\left[\frac{1}  {\w-H_{{\rm 
xx}}-\Sigma(\omega) +i\eta } \right],
\label{spec}
\ee
where the embedding self-energy reads $\Sigma(\omega)=
\sum_{{\l\l' \neq {\rm x }}} \d H_{{\rm x } \l} 
(\omega - H +i\eta)_{\l \l'}^{-1}
\d H_{\l' {\rm x } } $ and $\eta$ is a small broadending parameter.
A bound state in the continuum  with real energy  $\w_{0}$ exists if
$\w_{0}$ simultaneously satisfies the coupled self-consistency conditions
~\cite{PhysRevB.80.165125} $(i)$ ${\rm Im \S(\w_{0})}=0$ and  $(ii)$
$\w_{0}=H_{{\rm xx}}+{\rm Re \S(\w_{0})}$. When these criteria are met, the effective coupling 
$\d H$ becomes completely transparent to the renormalized 
A-exciton, indicating that this state is effectively 
decoupled from the e-h band.
In the Supplementary Note 4 we demonstrate that these conditions
are always satisfied under physically realistic
and experimentally relevant assumptions -- conditions that are entirely met in monolayer WS$_{2}$.

In Fig.~\ref{fig4} we present both $A_{{\rm x}}(\omega)$ 
and $\S(\w)$ for thermal and superfluid carrier populations at
a high excitation density of $n=10^{13}{\rm cm}^{-2}$.
For thermal populations (see Fig.~\ref{fig4}a), it is evident that the two bound-state 
conditions are never satisfied: the energy $\w_{0}$ at which the line $\omega -H_{{\rm xx}}$
crosses $\rm{Re}\S$ does not correspond with a zero of $\rm{Im}\S$.
A direct comparison with the equilibrium spectral function [i.e. when 
$\d H=0$ in Eq.~(\ref{spec})] confirms that the A-exciton ceases to exist as a bound state, 
as it decays through coupling to the renormalized continuum of modes.
In contrast, when the carriers follow a superfluid distribution (see 
Fig.~\ref{fig4}b), the energy
$\w_{0}=E_{{\rm x}}=\mu_{c}-\mu_{v}=2.15~$eV satisfies the bound-state criteria exactly: 
at this energy, $\omega -H_{{\rm xx}}$ intersects $\rm{Re}\S$ 
precisely where $\rm{Im}\S$ vanishes.
Consequently, the spectral function remains sharply peaked 
around $\w=\w_{0}=E_{{\rm x}}$, with a linewidth essentially identical to 
that of the equilibrium A-exciton.
This establishes that the exceptional exciton is effectively 
decoupled from the surrounding excitation continuum, allowing 
it to persist as a robust, spatially localized bound state. As
we show below, this localization gives rise to a set of striking optical 
and electronic properties that stem directly from the intrinsically 
non-radiative character of this exceptional eigenmode.

\begin{figure}[tbp]
  \centering
  \includegraphics[width=0.99\textwidth]{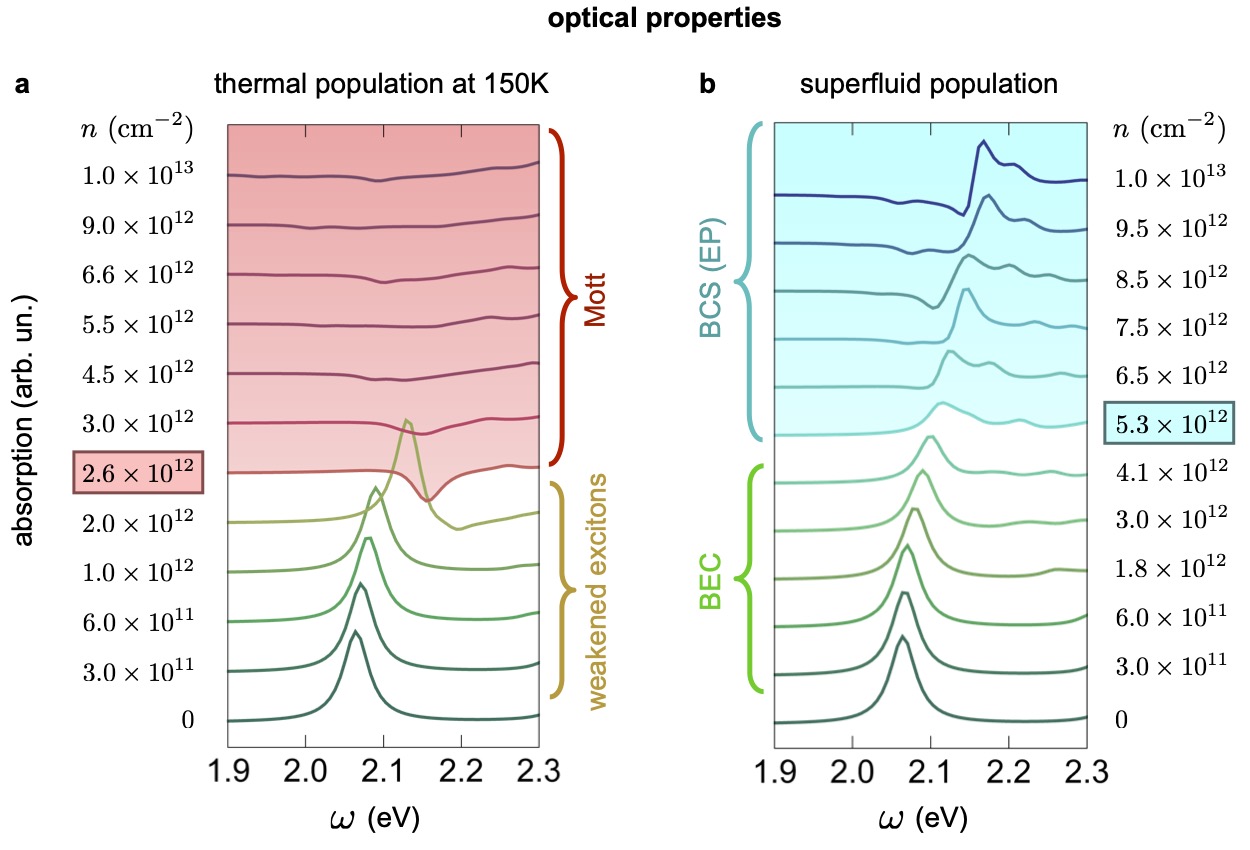}
  \caption{{\bf Optical properties.} Panel a: Calculated absorption spectrum of 
  WS$_{2}$ in the presence of thermal carriers at $T=150$~K  for different 
  excitation densities. The red-shaded region marks the onset and progression
  of the Mott regime, where bound excitons disappear.
  Panel b: Same as panel a, but for superfluid 
  carrier populations. The blue-shaded region denotes the regime in which exceptional
  excitons emerge (BCS regime).}
  \label{fig5}
  \end{figure}

\subsection*{Optical Absorption}
We now turn to the optical properties of 
the exceptional exciton. To highlight its peculiar behavior, 
it is instructive to first recall the characteristics of absorption 
spectra in thermally excited materials across the excitonic Mott transition.
Figure~\ref{fig5}a shows how the absorption spectrum of WS$_{2}$ evolves with increasing
excitation density, from equilibrium ($n = 0$) to the highly photoexcited
regime. When the renormalized $A$-exciton remains bound ($n < n_c$), the spectrum 
is dominated by a sharp absorption peak that progressively blue-shifts
as $n$ increases. At the Mott density, an abrupt spectral change occurs:
the excitonic peak suddenly vanishes and is replaced by a weak 
optical-gain region. For $n > n_c$, this gain region shifts toward
lower energies, consistent with a pronounced band-gap renormalization 
and population inversion. At higher energies, a region of weak positive
absorption persists, in agreement with previous theoretical~\cite{haug2009,PhysRevLett.84.2010,10.1063/1.5017069}
predictions and experimental~\cite{chernikov2015,PhysRevLett.134.066904}
observations.

When the carriers instead follow a superfluid distribution,
the absorption properties change dramatically, see Figure~\ref{fig5}b. At low excitation densities,
the behavior resembles that of the thermal case. However, beyond
the critical density where population inversion sets in, a distinct
excitonic feature survives and continues to dominate the spectrum.
In this regime, the absorption peak is preceded at slightly lower
energy by a strong gain anti-peak of comparable spectral weight. 
Remarkably, the absorption spectrum undergoes a sign reversal exactly at
$\omega = E_{{\rm x}} $, where it vanishes. 
This zero crossing is a direct manifestation of the exceptional-point
character of the excitonic mode.
At the critical energy $E_{{\rm x}}$, perfect destructive interference
between the gain and loss channels -- corresponding to the coalescence 
of complex eigenvalues with opposite imaginary parts -- leads to a complete balance of 
absorption and emission. The exciton thus becomes perfectly dark:
it ceases to radiate
and instead traps its energy indefinitely, corresponding to a theoretically
infinite quality factor 
$Q$~\cite{highq1,PhysRevLett.119.243901,PhysRevLett.121.193903,Bezus:18,PhysRevLett.134.103802,EBIC2}. 

\begin{figure}[tbp]
  \centering
  \includegraphics[width=0.99\textwidth]{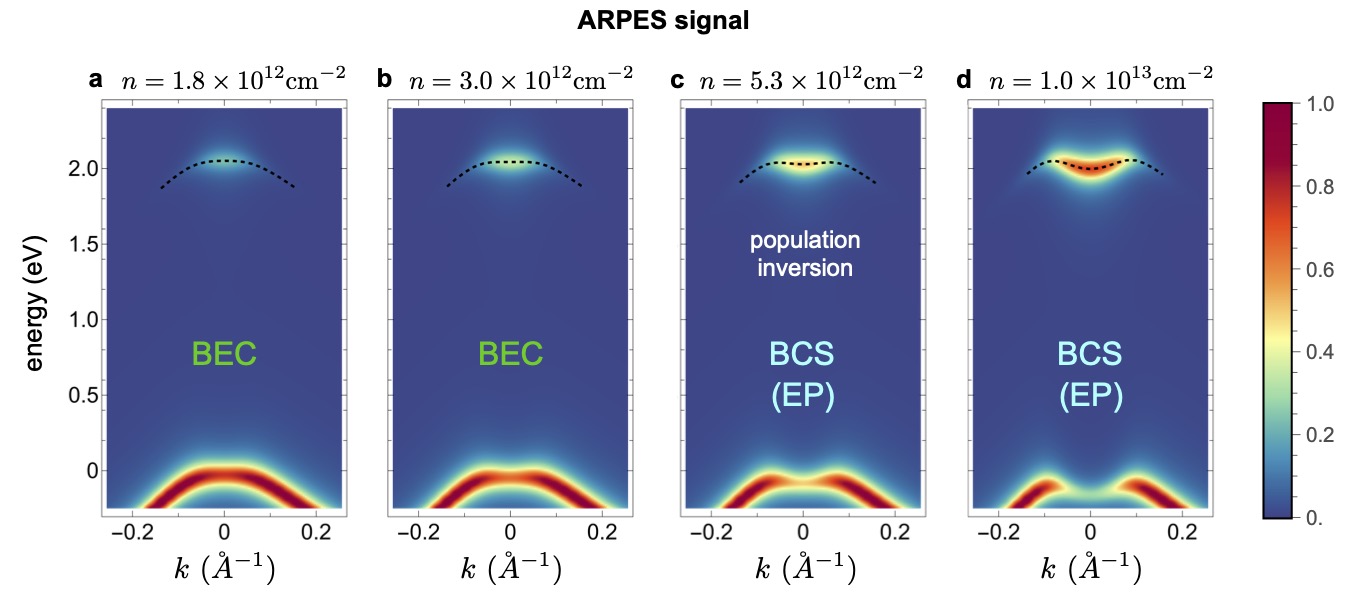}
  \caption{{\bf ARPES signal from exceptional excitons.}
  ARPES signal of 
  WS$_{2}$ computed  as $f^{{\rm sf}}_{v \blk}\delta(\w-E_{v 
  \blk})+f^{{\rm sf}}_{c \blk}\delta(\w-E_{c \blk})$  in the presence of superfluid carrier populations for different 
  excitation densities. Panels a and b correspond to the BEC regime, while panels c and d 
  illustrate the BCS regime, where exceptional excitons emerge. In all calculations, the delta functions 
  are approximated by Lorentzian profiles with a finite width of 70~meV.
  }
  \label{fig6}
  \end{figure}

\subsection*{ARPES Signal}
To identify the possible signatures of eigenvalue coalescence
in the ARPES spectrum, we compute the photoemission signal.
The ARPES intensity at energy $\w$ and momentum $\blk$ 
arising from electrons emitted from band $\nu$ is 
proportional to the lesser Green's function
$ G^{<}_{\nu \blk}(\w)$~\cite{svl-book,PhysRevB.94.245303}.
For resonantly excited 
semiconductors where the wavefunction of the A-exciton
has predominant contribution on a single conduction band $c$
and on a single valence band $v$, we have shown that 
$ G^{<}_{\nu 
\blk}(\w) \approx 2\pi i f^{{\rm sf}}_{\nu \blk}
\delta(\w-E_{\nu \blk})$~\cite{PSMS.2019,PBS.2020}.
The dispersion of the ARPES peak originating from a conduction photoelectron is given by
\be
E_{c \blk}=\frac{1}{2}\left[
\e^{{\rm r}}_{c \blk}+\e^{{\rm r}}_{v \blk}+E_{{\rm x }}-
\sqrt{(\e^{{\rm r}}_{c \blk}-\e^{{\rm r}}_{v \blk}-E_{{\rm x }})^{2}+4|\D_{\blk}|^{2}}
\right],
\label{arpes}
\ee
while the corresponding valence quantity reads  $E_{v \blk}=E_{c \blk}-E_{{\rm x }}$.
Here $\D_{\blk}$ denotes the $cv$ component of  $h^{\rm{HSEX}}_{\blk}$ in 
Eq.~(\ref{eq:exciton_sf}). Since the amplitude of $\D_{\blk}$ is proportional to the square root 
of the excitation density $\sqrt{n}$, the above expression predicts
that for low pumping the ARPES spectrum exhibits an excitonic 
sideband at energy $\omega = \e_{v \blk}+E^{{\rm eq}}_{{\rm x}}$.
This sideband corresponds to a replica of the valence band 
shifted by the A-exciton energy, as clearly visible in
Fig~{\ref{fig6}}a,b for densities 
$n<n_{c}$~\cite{PSMS.2019,PBS.2020,RustagiKemper2018}.
Such excitonic replicas have been observed in recent time-resolved ARPES
measurements in  different resonantly photoexcited 
materials~\cite{doi:10.1126/science.aba1029,doi:10.1126/sciadv.abg0192,https://doi.org/10.1002/ntls.10010,doi:10.1021/acs.nanolett.1c01839}.

As the excitation density increases, the ARPES signal undergoes a 
qualitative transformation.
For $n \geq n_{c}$ population inversion sets in and the excitonic 
eigenvalues coalesce. Concomitantly, the exciton energy
$E_{{\rm x }}$ enters 
in the e-h continuum and the term $\e^{{\rm r}}_{c \blk}-\e^{{\rm r}}_{v \blk}-E_{{\rm x }}$
in Eq~(\ref{arpes})
changes sign for momenta $\blk$ near to the $K$-point, signaling a 
strong band hybridization.
Around these momenta the excitonic sideband becomes heavily 
renormalized,
and $E_{c \blk}$ evolves into  a  replica of the 
conduction band, see Fig~{\ref{fig6}}c,d.
This feature signals the passage of the exciton superfluid through a BEC–BCS crossover and gives
rise to a characteristic Mexican-hat shape~\cite{PSMS.2019} .
Recent ARPES measurements on resonantly driven WS$_{2}$ 
monolayers have reported a very similar evolution of the 
spectrum~\cite{pareek2024driving}.
Our analysis 
reveals that the BEC-BCS crossover arises precisely
in correspondence of the formation of the exceptional excitons.

\section*{Discussion}
This work uncovers a previously unidentified mechanism by which
non-Hermitian systems can sustain real eigenenergies without
the need for finely tuned gain–loss balance. 
Here, non-Hermiticity arises naturally from the
asymmetric electron–hole interactions that develop under intense 
photoexcitation,
while real spectral branches emerge solely from intrinsic excitonic
correlations. Within this non-equilibrium landscape, we identify 
exceptional excitons as a new class 
of excitonic states. These represent 
exceptional points of the pseudo-Hermitian Bethe–Salpeter 
Hamiltonian that are embedded inside the electron–hole 
continuum, defying the conventional expectation that EP
require careful gain–loss engineering.
We demonstrate that exceptional excitons can be generated by
illuminating excitonic materials with appropriate pump pulses—conditions
that are readily accessible with current ultrafast laser technology. 
Crucially, the pump frequency and fluence must be tuned within an
optimal resonant window that supports
a superfluid-like phase with population inversion, enabling excitonic
eigenmode coalescence within the continuum.

We show that exceptional excitons remain isolated from the surrounding electron–hole
background through a subtle but robust self-organized balance
between gain and loss amplitudes. This mechanism suppresses
hybridization with other modes, rendering the resulting states 
particularly resilient to the excitonic Mott transition.
Our findings provide also a new perspective on the nature of non-equilibrium 
excitonic insulators at high photoexcited densities~\cite{PSMS.2019}. The formation of 
exceptional excitons constitutes an intrinsic, self-protecting mechanism 
for sustaining excitonic coherence in the BCS 
regime~\cite{PhysRevB.102.085203}.

We further predict distinct 
experimental signatures of exceptional excitons in both 
optical and photoemission observables. In linear absorption,
they appear as optically dark states whose energy marks the 
boundary between domains of strong gain and strong absorption.
In time-resolved ARPES, exceptional excitons emerging at high 
excitation densities are predicted to invert the curvature of
the excitonic sideband, producing a characteristic mexican-hat 
dispersion. The recent observation of such a feature in resonantly
driven WS$_{2}$ monolayers~\cite{pareek2024driving}
suggests that exceptional excitons
could have already been realized experimentally, pointing to a direct 
connection between eigenmode coalescence and the excitonic BEC–BCS crossover.

More broadly, the ability to generate real-energy eigenmodes
in non-Hermitian systems is highly desirable. 
In this regime, gain and loss channels conspire
to produce eigenstates with conserved total amplitude, ensuring
dynamical stability while retaining the exceptional sensitivity, 
topological structure, and spectral controllability inherent to non-Hermitian physics. 
In the specific case of exceptional excitons, 
this unique combination of stability and non-Hermitian functionality
opens promising avenues for practical applications requiring ultrahigh-Q 
resonances~\cite{PhysRevLett.134.103802,EBIC2}, including 
platforms for high-sensitivity sensors and low-loss nonlinear photonics.

\section*{Methods} \label{sec11}
\subsection*{First-principle modelling of WS$_{2}$ monolayers}
We employ the spin-dependent DFT band structure $\e_{\n \blk}$ parametrized 
within a tight-binding model as reported in  Ref.~\cite{PhysRevB.88.085433}.
To reproduce the experimental quasiparticle gap, 
the conduction bands $\e_{c \blk}$ are rigidly shifted upward by 
$0.8$~eV~\cite{PhysRevLett.113.076802,Hsu_2019}. 
The Bloch Hamiltonian eigenvectors ${\bf U}_{\blk }$
are used to construct the Coulomb matrix elements entering both the
Bethe–Salpeter kernel and the Hartree–SEX Hamiltonian.
These matrix elements are defined as~\cite{PhysRevB.91.075310}
\be
V_{imjn}^{\blq\blk \blk'}=v_{q} (\blU_{i \blk }^{\dag} \cdot \blU_{n \blk -\blq })   (\blU_{ 
m \blk ' }^{\dag} \cdot \blU_{j \blk ' +\blq })
\label{vrk}
\ee
and describe the scattering amplitude for two electrons initially 
in bands $j$ and $n$ with momenta
$\blk'+\blq$ and $\blk-\blq$ respectively to scatter into 
 bands $m$ and $i$ with 
momenta $\blk'$ and $\blk$. The interaction  $v_{q}$
corresponds to the Rytova–Keldysh potential~\cite{keldysh,PhysRevB.84.085406}
in momentum space i.e.
\be
v_{q}=\frac{2\pi}{ \ve q(1+r_{0}q)} ,
\ee 
where $q=|\blq|$, $r_{0}=22~\mathring{\rm A}$~\cite{Hsu_2019}
and $\ve=(\ve_{{\rm top}}+\ve_{{\rm bottom}})/2$ denotes the effective dielectric 
constant of the environment. In this work we use the representative values
$\ve_{{\rm top}}=1$ and $\ve_{{\rm bottom}}=3.1$ corresponding to 
a WS$_{2}$  monolayer on sapphire substrate and exposed to air~\cite{Hsu_2019}.
The  divergence of $v_{q}$ near the $\Gamma$-point is regularized by following 
Refs.~\cite{PhysRevB.88.245309,PhysRevB.97.205409}, i.e. 
by averaging $v_{q}$ on a very small domain $\Omega$ centered at
$\blq=0$, with linear size equal to the Brillouin-zone discretization step.
In semiconductors the Coulomb integrals that do not conserve
the number of valence and conduction electrons are typically very small
and can be neglected~\cite{PhysRevB.93.205145}. 
Furthermore, because the Rytova–Keldysh interaction tensor $V$ in Eq.~(\ref{vrk}) 
already captures the ab-initio static screening over the momentum 
range relevant for excitons,
the statically screened exchange interaction $W$ entering  the 
Bethe-Salpeter kernel and  the HSEX Hamiltonian can be  
reliably approximated by 
$W\approx V$.
This modelization has demonstrated remarkable accuracy 
in reproducing excitonic binding energies in 2D 
materials~\cite{PhysRevB.88.045318,PhysRevLett.113.076802,steinhoff2}.
For the present simulations, the active space is restricted to the two highest
valence and two lowest conduction bands. Accurate modeling of excitonic
effects in two-dimensional transition-metal dichalcogenides requires dense
sampling of the Brillouin zone; however, direct diagonalization 
of the non-Hermitian Bethe-Salpeter Hamiltonian in
Eq.~(\ref{phbsh}) and the self-consistent solution of
Eq.~(\ref{eq:exciton_sf}) in grids containing thousands of 
$\bf{k}$-points is computationally prohibitive.
As we focus on low-energy excitations in  WS$_{2}$, 
we limit our calculations to a small plaquette  $\mathcal{P}$
surrounding the $K$($K'$) valleys. The plaquette area $A$ and number 
of $\blk$-points $N_{\blk}$ therein
are chosen to satisfy two conditions: (i) the equilibrium 
Bethe-Salpeter spectrum obtained with
$\blk\in\callP$ matches the full-Brillouin-zone spectrum for 
energies $\w \lesssim 2.4~$eV; and (ii) all
states accessed by the carriers during the 
real-time dynamics under pumping at central photon energy
$\w_{{\rm P}} \lesssim 2.2~$eV (see Fig.~\ref{fig2}) lie within 
$\mathcal{P}$.
In the present study we  considered  $A=0.4~\AA \times 0.4~\AA$ and $N_{\blk}=729$,
corresponding to a $108\times 108$ grid for the entire  first Brillouin zone.

\subsection*{Photoexcited dynamics}
We detail here the methodology used to simulate the real-time dynamics of
WS$_{2}$ under excitation by a sub-gap pump pulse 
$\mathbf{E}_{{\rm P}}(t)$. The electronic dynamics are obtained by propagating
the one-particle density matrix $\r_{ij \blk}$, where $\blk$ denotes the cristal momentum,
$i,j$ are the spin-band indices that can be valence ($v$) or conduction 
$(c)$. Our approach is formulated within the Hartree plus statically screened exchange
(HSEX) approximation of many-body perturbation theory.
The time evolution is governed by the equation of motion~\cite{svl-book}
\bea
i\frac{d}{dt}\rho_{ij \blk }(t)&=&[h^{\rm{HSEX}}_\blk(t),\rho_\blk(t)]_{ij},
\label{eom}
\eea
where $h^{\rm HSEX}$ is the HSEX hamiltonian
given by
\be
h^{\rm{HSEX}}_{ij \blk }(t) 
= \d_{ij}\e_{i \blk }
+\sum_{\blk' mn}
(V_{imnj}^{\bf{0} \blk\blk'}-V_{imjn}^{(\blk-\blk') \blk 
\blk'})\d\r_{nm \blk' }(t) + \mathbf{E}_{{\rm P}}(t)\cdot 
\mathbf{d}_{ij \blk  }.
\label{hf}
\ee
Here $\d\r_{\blk}(t)=\r_{\blk}(t)-\r_{\blk}(0)\equiv \r_{\blk}(t) - 
\r_{\blk}^{{\rm eq}}$ 
denotes the deviation of the density matrix from its equilibrium 
value, while $\mathbf{d}_{\blk ij }$ are the interband dipole matrix
elements associated with optical transitions
from band $i$ to band $j$ defined as~\cite{perfetto_real-time_2023}
\be
\bld_{ij \blk }=\frac{1}{i}\frac{1}{\e_{i\blk }-\e_{j\blk }} \blU_{ i\blk }^{\dagger}\cdot 
\partial_{\blk}h_{\blk} \cdot  \blU_{j \blk }, 
\ee
where $h_{\blk}$ is the Bloch Hamiltonian.
From the time-dependent density matrix we extract the momentum-resolved
carrier populations in band $\nu$ as $f_{\blk \nu}(t) =\r_{\n \n \blk }(t)$.
The corresponding excitation density is defined as
$n=\frac{2}{N_{\blk}A}\sum_{\blk c} f_{c \blk }$, where $A=8.82~\AA^{2}$ is the 
area of the WS$_{2}$ unit cell; the prefactor of 2 accounts for 
the fact that the simulations explicitly include only the $K$ valley.
In the real-time calculations we initialize the system in the ground state of
WS$_{2}$, characterized by  $\r^{{\rm eq}}_{cc\blk}=\r^{{\rm eq}}_{cv\blk}=0$, and 
$\r^{{\rm eq}}_{vv\blk}=1$. The system is then driven by a pump pulse linearly polarized along the 
$x$-axis, with temporal duration of approximately 100~fs (see 
Fig.~\ref{fig2}).
The equation of motion in Eq.~(\ref{eom}) is integrated numerically
using a 4th order Runge-Kutta solver with 
a time step $\Delta t = 0.1$~fs.

\subsection*{Self-consistent calculation of superfluid populations}
When a semiconductor supporting a bright bound exciton is
resonantly photoexcited, the transient many-body state established after the pump pulse
corresponds to an excitonic superfluid, which is described 
by the BCS-like wavefunction~\cite{PSMS.2019,PBS.2020,PhysRevLett.125.106401}
 \be
 |\Phi \ket =\prod_{\blk \xi} 
 \left(\sum_{i=v,c}\varphi^{\xi}_{i \blk }
 \hat{d}^{\dag}_{i \blk }\right) |0\ket ,
 \ee
where the operator $ \hat{d}^{\dag}_{i \blk }$ creates an electron
with momentum $\blk$ in the band $i$, and the state $|0\ket$ denotes 
the electron vacuum. The coherence factors $\varphi^{\xi}$ 
are the eigenvectors of the 
self-consistent secular equation~\cite{PSMS.2019,PBS.2020,PhysRevLett.125.106401} 
\be
(h^{\rm{HSEX}}_{\blk} -\mu )\varphi^{\xi}_{\blk}=e^{\xi}_{\blk}\varphi^{\xi}_{\blk}
\label{sec}
\ee
and correspond  (at zero temperature) to the lower eigenvalues 
$e^{\xi}_{\blk}$ (see below).
In Eq.~(\ref{sec}) $ h^{\rm{HSEX}}_{\blk}$ denotes the  HSEX hamiltonian 
introduced in Eq.~(\ref{hf}) evaluated in the absence of the pump 
field ($\mathbf{E}_{{\rm P}}=0$). The $\blk$-independent diagonal
matrix $\mu$ incorporates distinct chemical 
potentials for valence and conduction electrons. Its matrix elements 
are
$\mu_{ij}=\d_{ij} \m_{i}$ with $\mu_{i}=\mu_{v}$ for valence and 
 $\mu_{i}=\mu_{c}$ for conduction. The explicit values of $\mu_{c}$ 
 and $\mu_{v}$ depend on the pump fluence, and determine the 
 excitation density.
The secular equation in Eq.~(\ref{sec}) must be solved 
self-consistently, as HSEX Hamiltonian depends 
on the eigenvectors $\varphi^{\xi}$ through 
\be
\r_{ij\blk }=\bra  \Phi |  
 \hat{d}^{\dag}_{j \blk }\hat{d}_{i \blk } |\Phi \ket
 =\sum_{\xi} \varphi^{\xi}_{i \blk } \varphi^{\xi 
 *}_{j \blk }.
\ee
In 2D transition metal dichalcogenides such as WS$_{2}$,
the A-exciton wavefunction is dominated by contributions from a 
single conduction band $c$ and a single valence band $v$. 
Moreover the A-exciton is tightly localized in the 
 $K (K')$ valleys, and consists predominantly of electrons in 
 $c\ua(\da)$ and holes in $v\ua(\da)$ respectively.
Owing to this strong band selectivity, the BCS equation describing the 
photoexcited A-excitonic insulator reduces to the 
 $2\times 2$ problem
\begin{align}
    \left(\!\!
\begin{array}{cc}
  \e^{{\rm r}}_{v\blk} - \m_{v}  &  
 \D_{\blk} \\ 
   \D^{*}_{\blk}& \e^{{\rm r}}_{c\blk} -\m_{c} \\ 
\end{array}\!\!
\right)\!\!
\left(\begin{array}{c}\varphi^{\pm}_{v\blk}\\\varphi^{\pm}_{c\blk}\end{array}\right)
=e^{\pm}_{\blk}\left(\begin{array}{c}\varphi^{\pm}_{v\blk}\\\varphi^{\pm}_{c\blk}\end{array}\right).
\label{mu}
\end{align}
The renormalized band energies $\e^{{\rm r}}_{i \blk}$ read
\be
\e^{{\rm r}}_{i \blk}=\e_{i \blk}+\sum_{\blk' m}
(V_{immi}^{\bf{0} \blk\blk'}-V_{imim}^{(\blk-\blk') \blk 
\blk'})\d\r_{mm\blk' }
\ee
where have used that only Coulomb integrals that conserve the
number of valence and conduction electrons are non-vanishing;
the inter-band potential $\D_{\blk}$, instead, takes the form
\be
\D_{\blk}=\sum_{\blk'}
(V_{vcvc}^{\bf{0} \blk\blk'}-V_{vccv}^{(\blk-\blk') \blk 
\blk'})\d\r_{vc\blk' },
\ee
with $\r_{\blk ij}= \varphi^{-}_{i\blk } \varphi^{- *}_{j\blk }$
and $\d \r = \r -\r^{{\rm eq}}$.
Here $\varphi^{-}$ denotes the eigenvector associated with the lower eigenvalue 
$e^{-}_{\blk}$ of the effective Hamiltonian
\be
e^{-}_{\blk}=\frac{1}{2}\left[
\e^{{\rm r}}_{c \blk}+\e^{{\rm r}}_{v \blk}-\mu_{c}-\mu_{v}-
\sqrt{(\e^{{\rm r}}_{c \blk}-\e^{{\rm r}}_{v \blk}-\mu_{c}+\mu_{v})^{2}+4|\D_{\blk}|^{2}}
\right].
\ee
At self-consistency, the superfluid population in band $\nu$ is given 
by $f^{{\rm sf}}_{\n \blk}=|\varphi^{-}_{\n \blk} |^{2}$.
In the numerical solution of Eq.~(\ref{mu}) we set  the maximum of the 
equilibrium valence band to zero, and vary the  chemical potentials 
symmetrically according to
$\mu_{v}=(E_{g}-\D\mu)/2$ and $\mu_{c}=(E_{g}+\D\mu)/2$.
The BCS equation does not admit a superfluid solution  ($\D_{\blk} \neq 
0$) for $\D \m<E^{{\rm 
eq}}_{{\rm x}}$. For $\D \m \gtrsim E^{{\rm eq}}_{{\rm x}} $ 
superfluid solutions emerge with a very small excitation 
density $n$,  corresponding to the BEC regime.  When $\D \m$ increased
well beyond $E^{{\rm eq}}_{{\rm x}}$,
the exciton density becomes large and the system enters a population-inverted
state characteristic of the BCS regime.



\section*{Supplementary information}
\subsection*{Supplementary Note 1: Reality of exciton energy for 
arbitray superfluid population}

The non-equilibrium Bethe–Salpeter Hamiltonian 
$H$ in Eq.~(\ref{phbsh})
\be
H_{cv \blk , c'v'\blk' }=(\e^{\rm r}_{c\blk }-\e^{\rm r}_{v \blk })
\d_{\blk \blk '}\d_{vv'}\d_{cc'} -(f_{v \blk}-f_{c \blk}) K_{cv 
\blk , c'v'\blk' },
\ee
contains the renormalized band energies
\be
\e^{\rm r}_{i \blk }=\e_{i \blk }+\sum_{\blk' m}
(V_{i mm i}^{\bf{0} \blk\blk'}-W_{i m i m}^{(\blk-\blk') \blk 
\blk'})(\r_{mm\blk' }-\r^{{\rm eq}}_{mm\blk' })
\ee
and the irreducible kernel
\be
K_{cv \blk , c'v'\blk' }=W_{c v' v c'}^{(\blk-\blk') \blk \blk'}
- V_{c v' c' v}^{\bf{0} \blk\blk'},
\ee
where $V$ is the bare Coulomb repulsion, while $W$ is 
the statically-screened exchange interaction.
The above expression for the PH-BSH is obtained under 
the assumption that in the presence of excited 
carriers both the density matrix and 
the HSEX hamiltionian retain a diagonal intra-valence and intra-conduction
structure~\cite{svl-book}, that is
\bea
\r_{vv'\blk}&=& \d_{vv'}\r_{vv\blk}=\d_{vv'}f_{v\blk} \nonumber \\
\r_{cc'\blk}&=& \d_{cc'}\r_{cc\blk} =\d_{cc'}f_{c\blk} \nonumber \\
h^{{\rm HSEX}}_{\blk vv'}&=&\d_{vv'}h^{{\rm HSEX}}_{\blk vv} = 
\d_{vv'}\e^{\rm r}_{v \blk } \nonumber \\
h^{{\rm HSEX}}_{\blk cc'}&=&\d_{cc'}h^{{\rm HSEX}}_{\blk cc}=\d_{cc'}\e^{\rm 
r}_{c \blk }.
\eea
Consequently, to demonstrate that the superfluid state
$\Psi^{{\rm sf}}_{cv\blk}=\sum_{\xi}\varphi^{\xi \ast}_{c \blk}\varphi^{\xi}_{v \blk}$
is an eigenvector of $H$, we must impose the same diagonal structure on the
matrix elements 
of $\r$ and $h^{{\rm HSEX}}$  when solving the BCS problem
$(h^{\rm{HSEX}}_{\blk} -\mu 
)\varphi^{\xi}_{\blk}=e^{\xi}_{\blk}\varphi^{\xi}_{\blk}$.
Accordingly, at self-consistency we have
\bea
\sum_{\xi} \varphi^{\xi}_{c \blk } \varphi^{\xi *}_{c' \blk } &=& 
\d_{cc'}\r_{cc\blk } \equiv \d_{cc'} f^{{\rm sf}}_{c \blk} \nonumber 
\\
\sum_{\xi} \varphi^{\xi}_{v \blk } \varphi^{\xi *}_{v' \blk } &=& 
\d_{vv'}\r_{vv\blk } \equiv \d_{vv'} f^{{\rm sf}}_{v \blk} \nonumber 
\\
\sum_{\xi} \varphi^{\xi}_{v \blk } \varphi^{\xi *}_{c \blk }  &=& 
\r_{vc\blk }=\r^{*}_{cv\blk }, 
\label{rhos}
\eea
where the sum over $\xi$ is restricted to the lower-energy eigestates 
of $h^{{\rm HSEX}}$.
The above relations imply that the following 
equations are simultaneoulsy satisfied for each $\blk$
\bea
(\e^{{\rm r}}_{v \blk}-\m_{v})\varphi^{\xi}_{v \blk} +\sum_{c'}h^{{\rm 
HSEX}}_{\blk vc'}\varphi^{\xi}_{c' \blk} &=& 
e^{\xi}_{\blk}\varphi^{\xi}_{v\blk} \nonumber \\
(\e^{{\rm r}}_{c \blk}-\m_{c})\varphi^{\xi}_{c \blk} +\sum_{v'}h^{{\rm 
HSEX}}_{\blk cv'}\varphi^{\xi}_{v' \blk} &=& 
e^{\xi}_{\blk}\varphi^{\xi}_{c\blk},
\label{exbcs}
\eea
where we recall that
\be
h^{{\rm HSEX}}_{\blk cv}=[h^{{\rm HSEX}}_{\blk vc}]^{*}=\sum_{v' c' 
\blk'} (V_{cv' c' v }^{\bf{0} \blk\blk'}-W_{c v' v c'}^{(\blk-\blk') \blk \blk'})
\r_{ c' v' \blk'}
\ee
By taking the complex conjugate of the first equation, 
by multiplying by
$\varphi^{\xi}_{c\blk}$ and by $\varphi^{\xi *}_{v\blk}$ the first and 
second equations respectively, and by summing over $\xi$, 
Eqs.~(\ref{exbcs}) become
\bea 
(\e^{{\rm r}}_{v \blk}-\m_{v})\sum_{\xi} \varphi^{\xi *}_{v\blk} \varphi^{\xi}_{c\blk}  
&+& \sum_{\xi}\sum_{v'' c' c'' 
\blk'}  \varphi^{\xi' *}_{c' \blk } \varphi^{\xi}_{c \blk } 
(V_{c'v'' c'' v }^{\bf{0} \blk\blk'}-W_{c' v'' v c''}^{(\blk-\blk') \blk \blk'})
\sum_{\xi'}  \varphi^{\xi' *}_{v'' \blk' } \varphi^{\xi'}_{c'' \blk' 
} \nonumber \\
&=& \sum_{\xi} \varphi^{\xi *}_{v\blk} \varphi^{\xi}_{c\blk} 
e^{\xi}_{\blk} \nonumber \\
(\e^{{\rm r}}_{c \blk}-\m_{c})\sum_{\xi} \varphi^{\xi *}_{v\blk} \varphi^{\xi}_{c\blk}  
&+& \sum_{\xi}\sum_{v'' v' c'' 
\blk'}  \varphi^{\xi' *}_{v' \blk } \varphi^{\xi}_{v \blk }  (V_{c 
v'' c'' v }^{\bf{0} \blk\blk'}-W_{c v''  v c'' }^{(\blk-\blk') \blk \blk'})
\sum_{\xi'}  \varphi^{\xi' *}_{v'' \blk' } \varphi^{\xi'}_{c'' \blk' 
} 
\nonumber \\
&=& \sum_{\xi} \varphi^{\xi *}_{v\blk} \varphi^{\xi}_{c\blk} 
e^{\xi}_{\blk} \nonumber \\
\eea
Subtracting the two equations, using Eqs.~(\ref{rhos}),
and  identifying the BCS state
$\Psi^{{\rm sf}}_{cv\blk}=\sum_{\xi} \varphi^{\xi *}_{v \blk}\varphi^{\xi}_{c \blk}$
we obtain
\be
(\e^{{\rm r}}_{c \blk}-\e^{{\rm r}}_{v \blk})\Psi^{{\rm sf}}_{cv\blk}-
(f^{{\rm sf}}_{v \blk}-f^{{\rm sf}}_{c \blk})\sum_{\blk' v'' c''}
(W_{c v'' v c''}^{(\blk-\blk') \blk \blk'}-V_{c v'' c'' v }^{\bf{0} 
\blk\blk'})\Psi^{{\rm sf}}_{c''v''\blk'}=(\mu_{c}-\mu_{v})\Psi^{{\rm 
sf}}_{cv\blk}.
\label{phbse}
\ee
This shows that  $\Psi^{{\rm sf}}$ is an eigenvector of PH-BSH  
with {\it real eigenvalue} $E_{{\rm x}}=\mu_{c}-\mu_{v}$ for {\it any 
superfluid population}.
Another remarkable property follows directly from Eq.~(\ref{phbse}):
in the limit of small excitation density $n$, corresponding to 
$f^{{\rm sf}}_{v \blk}\to 1$ and 
$f^{{\rm sf}}_{c \blk}\to 0$, the superfluid wavefunction naturally reduces
to the equilibrium exciton, i.e.  $ \Psi^{{\rm sf}} \to  \Psi^{{\rm 
eq}}_{{\rm x}} $.
Therfore $ \Psi^{{\rm sf}}$ can be interpreted as the adiabatically 
evolved exciton wavefunction in the superfluid excited state.

\subsection*{Supplementary Note 2: Finite-size effects}

\begin{figure}[tbp]
  \centering
  \includegraphics[width=0.99\textwidth]{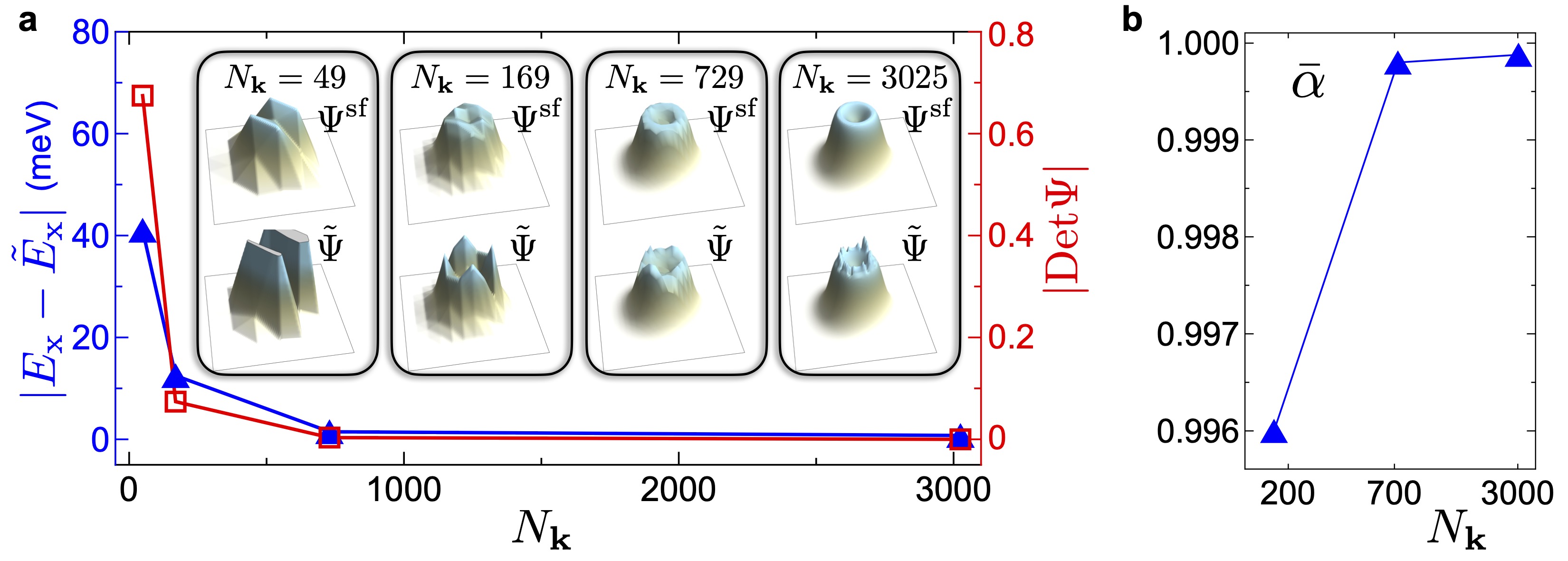}
  \caption{{\bf Finite-size effects.} Panel a: Energy splitting 
  $|E_{{\rm x}}-\tilde{E}_{{\rm x}}|$ (blue) and
  determinant of the eigevector matrix  $|{\rm Det} \Psi|$ 
  (red) as a function of number of $\blk$-points 
  $N_{\blk}$ in the square plaquette considered in all calculations.
  The insets display the rescaled eigenvectors
  $N_{\blk}|\Psi^{{\rm sf}}|^{2}$ and $N_{\blk}|\tilde{\Psi}|^{2}$
  for different values of $N_{\blk}$, demonstrating convergence of their spatial profiles.
  Panel b: Value of the control parameter $\bar{\a}$ at which 
  exact coalescence of $\Psi^{{\rm sf}}$  and $\tilde{\Psi}$ occurs
  for  fixed grid sizes $N_{\blk}=169,729,3025$. 
  }
  \label{fig7}
  \end{figure}

We have verified with high numerical precision that 
$\Psi^{{\rm sf}}$ is indeed an eigenvector of 
PH–BSH  when the latter is evaluated using 
the self-consistently determined superfluid populations
$f^{{\rm sf}}_{\nu \blk}$.
We further find that the exact realization of the exceptional point is
achieved only in the thermodynamic limit, i.e., when the number of 
$\blk$-points $N_{\blk}$ approaches infinity.
Figure~\ref{fig7}a illustrates the evolution of the eigenvector
$\Psi^{{\rm sf}}$ computed for a superfluid excitation density
$n=10^{13}~{\rm cm}^{-2}$ as a function of $N_{\blk}$.
The analysis of the spectrum reveals the presence of a second eigenvector,
$\tilde{\Psi}$, whose structure closely resembles that of
$\Psi^{{\rm sf}}$, and  whose associated eigenvalue $\tilde{E}_{{\rm x}}$
lies very close to $E_{{\rm x}}$. As $N_{\blk}$ increases, 
$\Psi^{{\rm sf}}$ and $\tilde{\Psi}$ converge rapidly toward 
one another, signaling the  eigenvector coalescence
characteristic of an exceptional point.
This increasing similarity is quantitatively reflected in the vanishing determinant
of the eigenvector matrix $|{\rm Det} \Psi|$ of PH-BSH.
Remarkably, already for $N_{\blk}=729$ (the value employed in the main calculations)
the coalescence of the two eigenvectors is nearly complete.
The corresponding eigenvalues also converge: the energy splitting
$|E_{{\rm x}}-\tilde{E}_{{\rm x}}|$ decreases monotonically with 
$N_{\blk}$, reaching approximately 1~meV at $N_{\blk}=729$.

It is important to emphasize that the behavior discussed above does not preclude
the exact emergence of exceptional excitons at finite $N_{\blk}$.
In such cases, eigenvalue and eigenvector coalescence can still 
be realized, but for carrier distributions
that deviate slightly from the fully self-consistent superfluid one.
Numerically, we find that for any finite $N_{\blk}$
and for any fixed excitation density $n$ there exists a value 
$\bar{\a}\lesssim 1$ for which a carrier distribution of the form
\be
f_{\nu \blk}=(1-\bar{\a})f^{\rm{th}}_{\nu \blk}+\bar{\a} f^{\rm{sf}}_{\nu \blk}
\ee
drives the system exactly to the excitonic exceptional point.
Here $f^{\rm{th}}_{\nu \blk}$ denotes the thermal distribution.
As illustrated in Fig.~\ref{fig7}b, for  $N_{\blk}=729$ and  an excitation density of
 $n=10^{13}~{\rm cm}^{-2}$, the exceptional point is reached for $\bar{\a}=0.9998$.
This indicates that a minimal contamination  with the thermal distribution is sufficient
to guarantee exact coalescence for any finite $N_{\blk}$.

\subsection*{Supplementary Note 3: Topological properties}

\begin{figure}[tbp]
  \centering
  \includegraphics[width=0.85\textwidth]{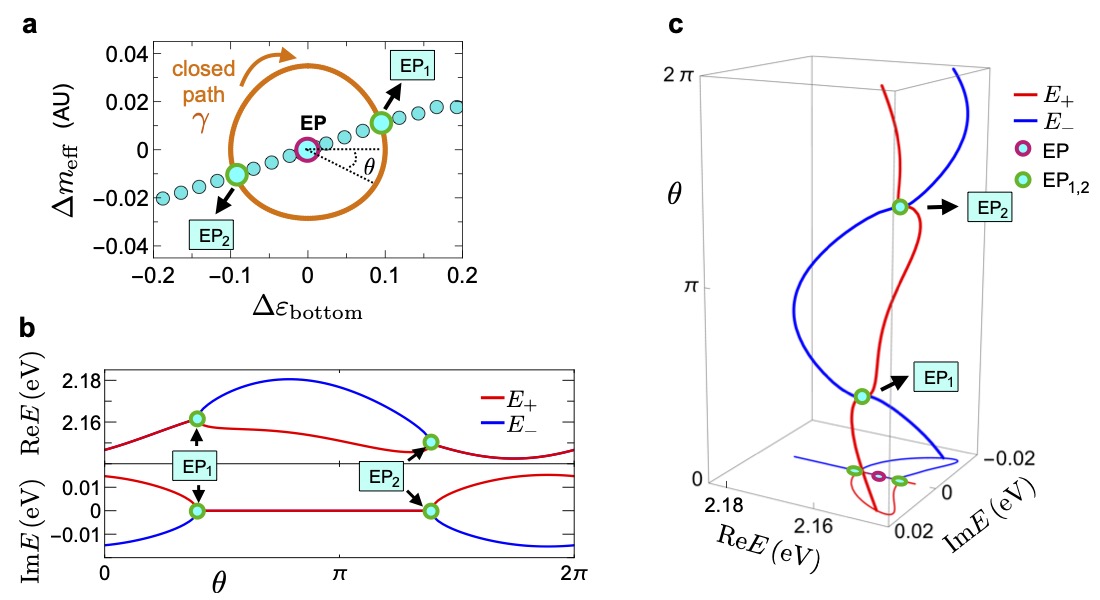}
  \caption{{\bf Topological properties.}   
  Panel a: Closed loop $\g$ encircling the excitonic exceptional point EP
  in the two-dimensional parameter space defined by the substrate dielectric constant 
 and the electron/hole effective masses.
 Here  $\D\ve_{{\rm bottom}}$ and $\D m_{{\rm eff}}$ denote deviations
 from the unperturbed values $ \ve_{{\rm bottom}} = 3.1$ and  
   $m^{c}_{{\rm eff}}=0.3~{\rm AU}$ and $m^{v}_{{\rm eff}}=0.4{\rm AU}$ 
   used throughout the main text. Turquoise circles mark the exceptional 
   line along which the exceptional exciton  resides. The two intersections between the path 
$\g$ and this line define the additional exceptional points EP$_{1}$ 
and  EP$_{2}$. Panel b: Real and imaginary parts of 
the eigenvalues $E_{\pm}$ as  functions of the winding angle $\theta$.
The additional exceptional points  EP$_{1}$ 
and  EP$_{2}$ are again highlighted by turquoise circles.
Panel c: Three-dimensional parametric trajectories of the real and imaginary 
parts of  $E_{+}$ and $E_{-}$ as a function 
   of the winding angle $\theta$. 
All calculations in this figure are performed at fixed
   supefluid  density $n=10^{13}~{\rm{cm}}^{-2}$.  
    }
  \label{fig8}
  \end{figure}

Exceptional points in non-Hermitian Hamiltonians are
known to generate a wide range of unconventional topological phenomena. 
In particular, second-order exceptional points possess a characteristic
topological invariant, called the {\it eigenvalue vorticity}, which captures the 
winding structure of the associated spectral Riemann surface.
Consider a non-Hermitian Hamiltonian $H(\bla)$ depending smoothly on a vector 
of control parameters $\bla$. Assume that at $\bla=\bla_{0}$ the 
spectrum exhibits an EP. Away from the EP, the defective eigenvalue 
splits into a pair of branches $E_{\pm}(\bla)$ of complex
eigenvales. The eigenvalue vorticity is defined as~\cite{EP3}
\be
\nu=\frac{1}{2\pi i} \oint_{\gamma} \grad_{\bla} 
\ln[E_{+}(\bla)-E_{-}(\bla) ] d\bla,
\label{vortex}
\ee
where $\gamma$ is a closed loop in parameter space encircling 
$\bla_{0}$.
A non-zero value of $\nu$ signals a non-trivial topological structure
of the spectral Riemann surface: it counts the
number of times one eigenvalue sheet winds around the other as
$\bla$ is transported adiabatically along $\gamma$. For example,
$\nu=1$ implies that a single traversal of the loop exchanges the two eigenvalue branches,
$E_{+}(\bla) \leftrightarrow E_{-}(\bla)$,
reflecting the characteristic square-root topology of a second-order 
EP.
Notice that evaluating the integral in Eq.~(\ref{vortex}) requires parallel transport, ensuring that both
$E_{\pm}(\bla)$ and the corresponding eigenvectors $\Psi_{\pm}(\bla)$ 
are tracked continuously along $\g$. This adiabatic continuation is essential for capturing the correct branch 
structure and thus the correct topological invariant.

To probe the topological properties of exceptional excitons,
we introduced a two-parameter deformation of the PH-BSH.
Fixing the superfluid density at  $n=10^{13}~{\rm{cm}}^{-2}$,
we varied (i) the substrate dielectric constant $\ve_{{\rm 
bottom}}$ appearing in the Rytova-Keldysh potential,
and (ii) the effective mass  $m_{{\rm eff}}$ that governs the 
band dispersions  
$\e_{\nu \blk}$  of conduction and 
valence bands. The corrsponding deviations $\D\ve_{{\rm bottom}}$
and $\D m_{{\rm eff}}$ trace the closed trajectory $\g$ shown in Fig.~\ref{fig8}a.
The exceptional exciton resides at the center of this parameter space  
$\D\ve_{{\rm 
bottom}}=\D m_{{\rm eff}}=0$, with exceptional energy $E_{{\rm 
x}}=2.15$~eV. Away from the EP,  $E_{{\rm 
x}}$ splits  into the complex-conjugate pair  $E_{\pm}$,
whose real and imaginary parts 
are displayed in  Fig.~\ref{fig8}b, as a function of the winding angle
$\theta={\rm Arg}[\D m_{{\rm eff}}/\D\ve_{{\rm bottom}}]$ (with $\D 
m_{{\rm eff}}$ expressed in atomic units). Remarkably, along this loop the 
eigenvalues become degenerate at two distinct values of $\theta$,
revealing the emergence of two additional exceptional points 
EP$_{1}$ and EP$_{2}$.
A closer inspection shows that the exceptional exciton is
not an isolated EP: it belongs to an exceptional line  
(turquoise circles in Fig.~\ref{fig8}a). Consequently, the eigenvalue vorticity
$\nu$ becomes ill-defined, because parallel transport along $\g$
is obstructed by eigenvector coalescence at EP$_{1}$ and EP$_{2}$.
We argue that this non-trivial scenario arises because the PH-BSH 
sits precisely on a critical manifold
where the topological invariant undergoes a discontinuous change.
Indeed the three-dimensional trajectrories of $E_{+}$ and $E_{-}$
displayed in Fig.~\ref{fig8}c closely mirror the behavior reported 
in Ref~\cite{PhysRevB.110.045444}, where the double intersection of
$E_{+}$ and $E_{-}$ along the loop signaled that the system was located 
exactly at the phase boundary between regions with vorticity 
$\nu=1$ and $\nu=-1$. 
Introducing an infinitesimal third parameter would lift this critical degeneracy,
regularize the branch structure, 
and restore a well-defined topological invariant of either $+1$ or 
$-1$.
  
\subsection*{Supplementary Note 4: On the bound-state condition}
  
To determine whether the excitonic exceptional point persists as a bound state in the continuum,
we analyze the singular behavior of the spectral function  
\be
A_{{\rm x}}(\omega)=-{\rm Im}[G_{{\rm x}}(\w)],
\label{specf}
\ee
where the embedded-exciton Green’s function is
\be
G_{{\rm x}}(\w)=\frac{1}  {\w-H_{{\rm xx}}-\Sigma(\omega) +i\eta }.
\ee
The renormalized energy  $H_{{\rm xx}}$ and the self-energy  $\S(\w)$ 
are defined in the main text.
Here $\Psi^{{\rm eq}}_{{\rm x}}$ denotes the bound-exciton wavefunction 
evaluated in equilibrium (i.e. for $n=0$),
which satisfies the equilibrium Bethe-Salpeter equation  
$H^{{\rm eq}}\Psi^{{\rm eq}}_{{\rm x}}=E^{{\rm eq}}_{{\rm 
x}}\Psi^{{\rm eq}}$.
As the superfluid excitation density  $n$
increases,  $\Psi^{{\rm eq}}_{{\rm x}}$ evolves adiabatically into 
$\Psi^{{\rm sf}}$, while the eigenvalue $E^{{\rm eq}}_{{\rm x}}$ 
evolves into $E_{{\rm x}}$. We recall, however, that only for
$n$ larger than a critical value $n_{c}$, $E_{{\rm x}}$ enters in 
the continuum and the exceptional exciton form.

To prove that for $n>n_{c}$ the exceptional-point energy
$E_{{\rm x}}$  
fullfills  the two bound-state conditions 
 $(i)$ ${\rm Im} \S(E_{{\rm x}})=0$ and  $(ii)$
$E_{{\rm x}}=H_{{\rm xx}}+{\rm Re} \S(E_{{\rm x}})$,
it is convenient to express $G_{{\rm x}}(\w)$ equivalently
as~\cite{grosso2013solid}
\be
G_{{\rm x}}(\w) =\left[\frac{1}{\w -H +i\eta} \right]_{{\rm xx}},
\ee
where the subscript ${\rm xx}$ denotes the matrix element over the 
state $\Psi^{{\rm eq}}_{{\rm x}}$.
In this form, satisfying the bound-state conditions is equivalent to showing that 
$G_{{\rm x}}(\w)$ diverges as $\eta^{-1}$ in the limit $\w\to E_{{\rm 
x}}$. A sufficient condition for such a divergence is that the overlap
$\bra \Psi^{{\rm eq}}_{{\rm x}} |  \Psi^{{\rm sf}} \ket  $ remains finite 
for any  superfluid density $n$. This ensures that the exceptional exciton wavefunction
$ \Psi^{{\rm sf}}$ is normalizable and spatially localized,
even though its energy lies within the continuum. Physically, this 
also implies that  $ \Psi^{{\rm sf}}$ does not couple to extended
states and therefore cannot decay into them,
preventing it from turning into a resonance.
The finitess of the overlap $\bra \Psi^{{\rm eq}}_{{\rm x}} |  \Psi^{{\rm sf}} \ket$
is justified on  physical grounds.
For any value of $n$, the eigenvector  $ \Psi^{{\rm sf}}$ represents the adiabatic
continuation of the equilibrium exciton wavefunction $\Psi^{{\rm 
eq}}_{{\rm x}}$ under a smooth  change of the populations
$f^{{\rm sf}}_{\nu \blk}$. 
In particular both $\Psi^{{\rm eq}}_{{\rm x}}$ and $ \Psi^{{\rm sf}}$ 
satisfy the same Bethe-Salpeter equation~\ref{phbse}, 
albeit with different occupation numbers. 
Under the reasonable assumption that the excitonic superfluid preserves the symmetry of the 
original exciton, adiabatic continuity in Hilbert space naturally 
prevents
 $ \Psi^{{\rm sf}}$ to become orthogonal to $\Psi^{{\rm eq}}_{{\rm 
 x}}$. 

 \begin{figure}[tbp]
  \centering
  \includegraphics[width=0.99\textwidth]{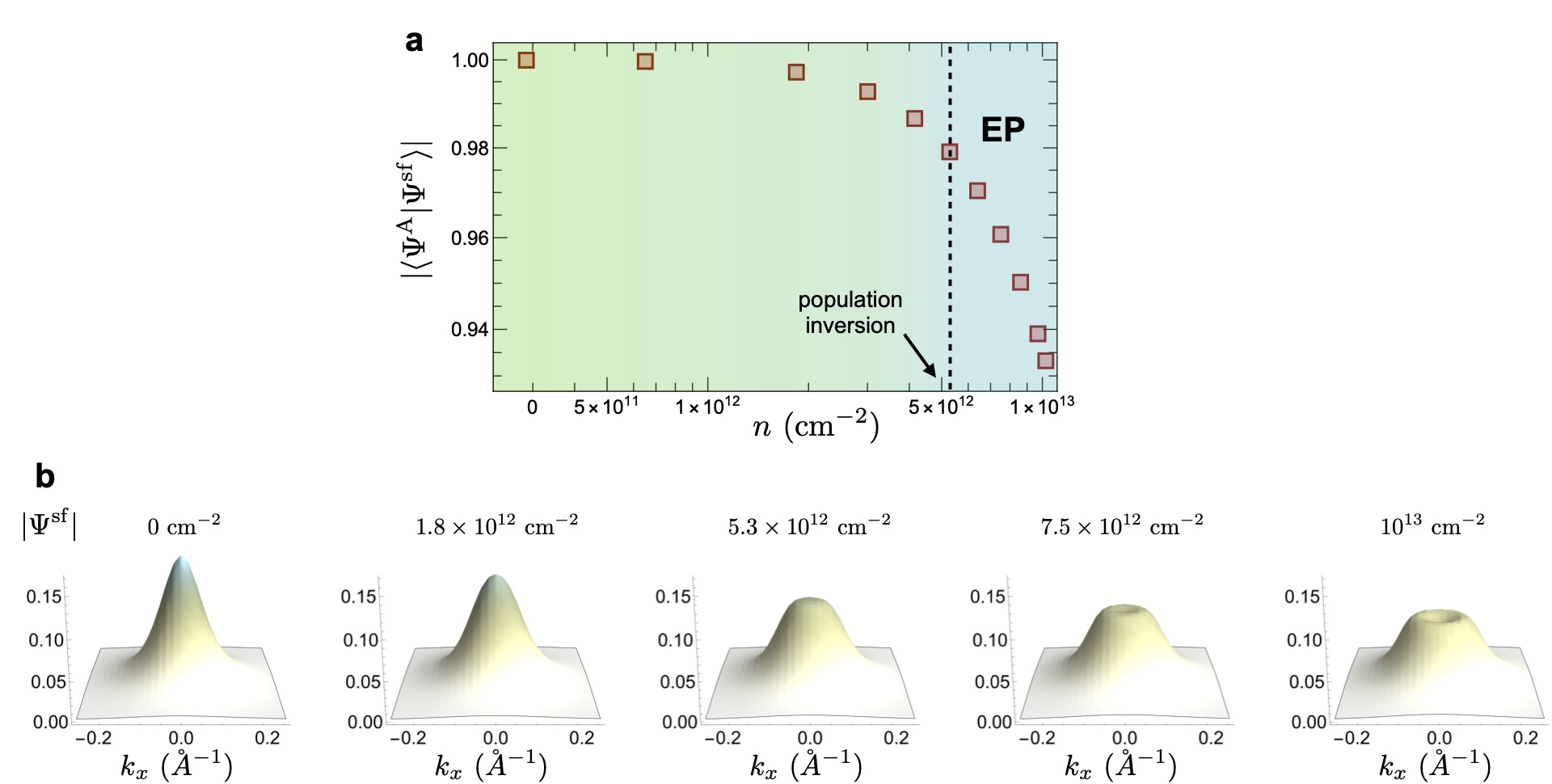}
  \caption{{\bf Bound-state condition.} Panel a:
  Absolute value of the overlap $|\bra \Psi^{{\rm eq}}_{{\rm x}} |  
  \Psi^{{\rm sf}} \ket|=|\bra \Psi^{{\rm A}} |  
  \Psi^{{\rm sf}} \ket|$ 
  as a function of the superfluid excitation density $n$.
  The dashed vertical line denotes the critical density $n_{c}= 5.3 \times 10^{12}~{\rm cm}^{-2} $
  above which exceptional excitons form.
  Panel b: Evolution of the superfluid exciton wavefunction $| \Psi^{{\rm sf}} |$
  for different values of  $n$. For $n\geq n_{c}$ $ \Psi^{{\rm sf}} 
  $ becomes the exceptional exciton wavefunction.
  }
  \label{fig9}
  \end{figure}

The non-orthogonality condition is unequivocally fulfilled in monolayer
WS$_{2}$. In this case
$\Psi^{{\rm eq}}_{{\rm x}} \equiv \Psi^{{\rm A}}$ 
is the A-exciton wavefunction, and the overlap 
$\bra \Psi^{{\rm eq}}_{{\rm x}} |  \Psi^{{\rm sf}} \ket$,
shown in Fig.~\ref{fig9}a as a function of the superfluid excitation density,
remains remarkably close to unity even up to the high-density regime 
 $n\approx 10^{13}~{\rm cm}^{-2}$.
 The robustness of this overlap is readily understood by examining the evolution of
the eigenvector $\Psi^{{\rm sf}}$ in Fig.~\ref{fig9}b.
Below the population inversion threshold  $n_{c}\approx 5.3 \times 10^{12}~{\rm cm}^{-2} $
the superfluid wavefunction is very similar to $\Psi^{{\rm A}}$,
featuring the same pronounced maximum at  $\blk=K$.
Above population inversion, where the exceptional exciton penetrates 
the electron–hole continuum,
$\Psi^{{\rm sf}}$ undergoes a qualitative restructuring: a local minimum develops at
$\blk=K$, encircled by a corona of enhanced weight.
Crucially, this deformation does not diminish its spatial localization, 
which remains comparable to that of the equilibrium A-exciton.
This persistent localization is a direct signature of the  bound-state 
character of the exceptional exciton, and 
explains the consistently large overlap observed in Fig.~\ref{fig9}a

\section*{Data availability}

The data supporting the findings of this study are available upon request.

\section*{Acknowledgements}

We acknowledge funding from Ministero Università e Ricerca PRIN 
under grant agreement No. 2022WZ8LME, from INFN through project TIME2QUEST, 
from European Research Council
MSCA-ITN TIMES under grant agreement 101118915, and from 
Tor Vergata University through project TESLA.

\section*{Competing interests}

The authors declare no competing interests.


\vspace{1cm}
\begin{center}
\large
{\bf References}
\end{center}


%

\end{document}